# The 25 Parsec Local White Dwarf Population

J. B. Holberg[1], T. D. Oswalt[2], E. M. Sion[3], G. P. McCook[3]

[1] *Lunar and Planetary Laboratory, University of Arizona, Tucson, AZ 85721, USA*
[2] *Embry-Riddle Aeronautical University, Daytona Beach, FL. 32114, USA*
[3] Department of Astronomy and Astrophysics, Villanova University, 800 Lancaster Ave. Villanova
 *University, Villanova, PA, 19085, USA*

**ABSTRACT**

We have extended our detailed survey of the local white dwarf population from 20 pc to 25 pc, effectively doubling the sample volume, which now includes 232 stars. In the process new stars within 20 pc have been added, a more uniform set of distance estimates as well as improved spectral and binary classifications are available. The present 25 pc sample is estimated to be about 68% complete (the corresponding 20 pc sample is now 86% complete). The space density of white dwarfs is unchanged at $4.8 \pm 0.5 \times 10^{-3}$ pc$^{-3}$. This new study includes a white dwarf mass distribution and luminosity function based on the 232 stars in the 25 pc sample. We find a significant excess of single stars over systems containing one or more companions (74% vs 26%). This suggests mechanisms that result in the loss of companions during binary system evolution. In addition this updated sample exhibits a pronounced deficiency of nearby "Sirius-Like" systems. Eleven such systems were found within the 20 pc volume vs, only one additional system found in the volume between 20 pc and 25 pc. An estimate of white dwarf birth rates during the last ~ 8 Gyr is derived from individual remnant cooling ages. A discussion of likely ways new members of the local sample may be found is provided.

*E-mail: holberg@argus.lpl.arizona.edu

# 1 INTRODUCTION

There is considerable interest in establishing a complete census of the population of nearby stars, particularly for those of the lowest luminosity, such as white dwarfs (WDs), late M-stars and sub-stellar L and T dwarfs. For the WD stars, previous publications (Holberg et al. 2002, and 2008, hereafter LWD02 and LWD08, respectively) have discussed the sample extending to 20 pc from the Sun. Since 2008 the 20 pc sample has grown by 10 members and distance estimates have been improved for many other WDs in this range. In addition, Giammichele, Bergeron & Dufour (2012 - hereafter GBD) have conducted a thorough spectral analysis of most WDs within the existing 20 pc sample. These developments provide a firm basis for a more homogeneous and unbiased determination of spectral types and stellar parameters. Thus, an improved knowledge base now exists with which to characterize the 20 pc local sample, since its completeness now approaches 86%. In this paper our previous 20 pc (LS20) sample is updated and extended out to a distance of 25 pc (LS25). Our formal sample limit is actually 25.2 pc, since at this distance the added volume is exactly twice that of the original LS20 sample, which facilitates direct comparisons between the two subsamples. For example, the higher completeness factor of LS20 permits extrapolation of population properties to the less complete LS25.

The LS25 is a volume-limited sample in which distance is the fundamental criterion for inclusion. Such a sample has several properties that are not easily matched by magnitude-limited samples that include many more WDs. For example, LS25 is an all-sky survey that includes all known spectroscopically identified WDs regardless of temperature, magnitude or spectral type. Most other samples and surveys contain explicit limitations on spatial and/or angular coverage, apparent magnitudes, colors, proper motions and spectral types. These limits translate into implicit filters with respect to sky coverage, effective temperature, and absolute magnitude. Binary status is another key factor that is often not considered in other surveys.

This paper closely follows the discussion of the LS20 in LWD08. Specifically, the spatial homogeneity of the 25 pc sample is evaluated with respect to its centroid and its north-south asymmetry. In addition, the sample completeness is evaluated as a function of Galactic latitude, and of WD luminosity. We also predict the types of degenerate stars that remain to be found in the 25 pc and sample discuss how best to find them. Spectroscopic and/or photometric data now exists for most stars with sufficient precision to determine such parameters as mass, absolute luminosities, and cooling age. In a related paper Sion et al. (2014) discussed LS25 in terms of space motion and the distribution of WD spectral types.

Section 2 introduces the LS25 sample, which encompasses all known WDs out to a distance of 25.2 pc, including published lists of newly recognized WDs. This section provides the identities, spectral types, V magnitudes, and trigonometric parallaxes, where they exist. Also included are the nearby WDs known to be in binary/multiple star systems. Section 3 discusses the spatial and spectral distribution of the local sample together with its completeness as a function of distance. Section 4 presents analyses of the LS25 sample including the WD mass distribution, the nearby luminosity function, the distribution of WD cooling ages, as well as comparisons to prior published results. The fraction of the sample that includes components of binary and multiple star systems is also considered. Section 5 presents the prospects for extensions and improvements to the local sample. Section 6 summarizes the conclusions of this study.

## 2. THE LOCAL WHITE DWARF SAMPLE

Table 1 presents the basic properties of the 232 WDs in the LS25 identified by WD number and alternate name (Columns 1 - 2). If the star is not in the current on-line version of the *Villanova Catalog of Spectroscopically identified White Dwarfs*[1] (McCook & Sion 1999, hereafter MS99) a preliminary WD number is computed. Column 3 gives the adopted spectral type for each star following the MS99 conventions and adjusting the numerical temperature index in each spectral type to match our adopted temperature. Column 4 lists an observed V-band magnitude for each star. Where no V-band photometry exists, effective V magnitudes are calculated from alternate photometric data and the spectroscopic $T_{eff}$ and log g estimates. The measured trigonometric parallax and uncertainty in milliarcseconds (mas) and references (Columns 5, 6 and 7) are given for the stars, if such measurements exist. The three primary sources for these parallax data are the Yale Parallax Catalog (van Altena, Lee & Hoffleit 1994), the revised *Hipparchos* Catalog (van Leeuwen 2007) and several publications of *The Research Consortium on Nearby Stars*[2] (RECONS). Weighted mean parallaxes from Subasavage[3] (2013) are also used. In some cases the listed parallaxes are those of the common proper motion companion star. Column 8 indicates whether the WD is single (s) or a member of a binary or multiple star system. Distinctions are made between binary systems (b), multiple star systems (m), Sirius-Like systems (sl), and double degenerate systems (dd). As in LWD08, unresolved double degenerate systems are counted twice. Columns 9 and 10 list the number of unresolved WD components and comments (a) in Appendix A. Additional kinematic data (positions, proper motions, and radial velocities) for these stars is contained in Sion et al (2014).

*2.1 New Sources of Local WDs*

Since LWD08 and Sion et al. (2014) several studies and surveys have yielded new local 20 pc WD candidates as well as WDs within 25 pc. Holberg (2015) estimated the distances to 12,400 DA stars in the SDSS DR7 spectroscopic catalog of Kleinman et al. (2013). However, due to the SDSS bright limits, this large sample contained only *three* stars were estimated to be within 25 pc: WD 0805+356, WD 1148+687, and WD 2246+223. Of these, WD 2246+223 was previously known. In their examination of Sirius-Like Systems Holberg et al. (2013) included two recently discovered white dwarfs in systems: WD 0210-510 in GJ 86AB and WD 0416-593 in HD 27442 AB not included in the 20 pc sample of Holberg et al. (2008), but which were included in Sion et al. (2009). The WDs in these systems were discovered as close companions to main sequence stars during exoplanet studies of these systems. Farihi et al. (2013) obtained *HST* STIS spectra of the WD in GJ 86 demonstrating that it is a DQ star similar to Procyon B. Farihi et al. (2011) also obtained VLT spectra of the DA star HD 27442 B, which successfully established its $T_{eff}$ and log g.

---

[1] http://www.astronomy.villanova.edu/WDcatalog/

[2] http://www.recons.org/

[3] http://www.denseproject.com/

Limoges, Lépine, & Bergeron (2013) published a list of proper motion selected WDs whose spectroscopic/photometric distances placed them within 40 pc; 11 of these stars have estimated distances within 25 pc[4]. The nearest of these (WD 1630+089), at 13.3 pc, was also included in LWD08, Sayers et al. (2012) and Sion et al. (2009). The remaining 10 stars from Limoges et al. have been included in our current sample. Recently, Limoges, Bergeron & Lépine (2015, hereafter LBL) have extended the Limoges et al. (2013) results to include an additional 14 WDs within 25 pc. These two studies have significantly increased the N-S asymmetry of our sample, since their stars are exclusively from the northern hemisphere (see section 3.1). Secondly, the gravities (and hence masses) for these stars are systematically larger than the general WD population. Specifically, the mean mass for the 10 stars is 0.9 $M_\odot$, while that of the general WD population is 0.67 $M_\odot$ (Kleinman et al. 2013). This is an interesting result because the authors have included empirical corrections to account for the so called "high log g problem" (Tremblay et al. 2013) exhibited by DA stars between effective temperatures of 7000 and 14000 K. As pointed out by LBL, these larger masses may well be a consequence of lower luminosities that have been reached in their proper motion based survey relative to prior WD searches. Sayers et al. (2012) presented spectroscopic distances for 26 proper motion selected WDs, finding three within 25 pc: WD 1338+052[5], WD 1630+089, and WD 2119+040. These have been included in our present sample. Bergeron et al. (2011) conducted a comprehensive analysis of 108 DB stars in that included spectroscopic distance estimates. Three of these stars have spectroscopic distances that seem to place them within our 25 pc sample. However, it is likely that all are well beyond 25 pc, as their spectroscopic is log g values are biased towards large values (log g > 9). These stars are individually discussed in Section 4.5 on DB stars.

Kilic et al. (2010) studied a sample of 126 faint cool and ultra-cool WDs in the SDSS spectroscopic survey that were selected on the basis of high proper motions. Their estimated distances for these stars included three that are potentially within 25 pc;

---

[4] WD 0145+360, WD 0252+497, WD 0340+198, WD 0454+620, WD 0649+639, WD 0744+112, WD 1630,089, WD 1911+536, WD 1912+143, WD 2111+072

[5] Sayers et al. incorrectly lists this star as WD 1336+052 in their Table 1.

(1) *WD 0749+42:* The distance to this DC 9 star (J0753+4230) is estimated to be 24 pc. However, Kawka & Vennes (2006) give a photometric distance of 18 pc. Lacking a reliable surface gravity determination, its distance uncertainty is significant. This star is included in our LS25 sample.

(2) *WD 1401+457:* The distance to this cool DC star is also estimated to be 24 pc. However, the large temperature uncertainty (2670 ± 1500 K) and lack of a reliable surface gravity determination yields a very considerable distance uncertainty. This star is included in our sample based on the assumed surface gravity of log g = 8.0 ± 0.5 (d = 24 ± 12 pc).

(3) *WD 2322+137:* The distance to this DA 9 star (J2325+1403) is given as 15 pc, placing it in the 20 pc sample. However, Lépine et al. (2009) published a parallax placing the star at 22.3 pc.

All of these stars are included in our LS25 sample with appropriate distance uncertainties. Finally, Kawka & Vennes (2012) published a study of 58 cool DA stars, some of which contained distance estimates that are potentially within 25 pc. Discounting those already contained in the LS20 sample only one, WD 1145-451, is included in our LS25 study.

*2.2 Photometry*

Existing multi-band photometry for each star in our LS25 sample is listed in Table 2. This includes the UBVRI data from various sources in the literature. In Table 2 the $JHK_s$ photometry is primarily from the 2MASS *All-Sky Point Source Catalog*[6] (*PSC*). For those stars included in the Sloan Digital Sky Survey DR9 *SkyServer*[7] *ugriz* photometry is used but not reported here. When possible, Table 2 lists appropriate photometric estimated uncertainties. These values are often critically important in estimating the distance uncertainties. The UBVRI photometry we used is quite heterogeneous, and was taken primarily from the MS99. When multiple independent photometric measurements exist mean magnitudes and uncertainties were estimated. We used, where available, corrected *y*-band Strömgren photometry to estimate V-band magnitudes (see Holberg & Bergeron 2006). The primary use of the photometric data is to estimate distances and the corresponding uncertainties, either as a substitute for non-existent trigonometric parallaxes or to compare with trigonometric parallaxes as discussed in Holberg, Bergeron & Gianninas (2008, hereafter HBG.

**3. POPULATION CHARACTERISTICS AND PHYSICAL PROPERTIES**

*3.1 Spatial Distribution*

---

[6] http://irsa.ipac.caltech.edu/cgi-bin/Gator/nph-dd?catalog=fp_psc

[7] http://skyserver.sdss3.org/dr9/en/tools/chart/

Among the expected characteristics of a relatively nearby volume-limited stellar distribution are spatial homogeneity and areal isotropy. Significant departures from these expectations may indicate where additional local WDs are likely to be found. In Fig. 1 the celestial locations of the LS25 sample are displayed on an equal-area Hammer–Aitoff projection. The sample in LS08 exhibited a modest 5:4 N-S asymmetry. In the present LS25 sample this increases to a 4:3 N-S asymmetry (131:101). The primary cause of this increase is the new WDs identified by Limoges et al. (2011; 2015) which were identified from exclusively northern hemisphere proper motion surveys. The probability of finding and excess of 31 stars in the Northern Hemisphere due to statistical fluctuations is 1/133. For the present LS25 sample the computed centroid of the sample gives a location of: $\alpha = 20.2°$ and $\delta = +31.6°$ with an offset distance of 2.36 pc.

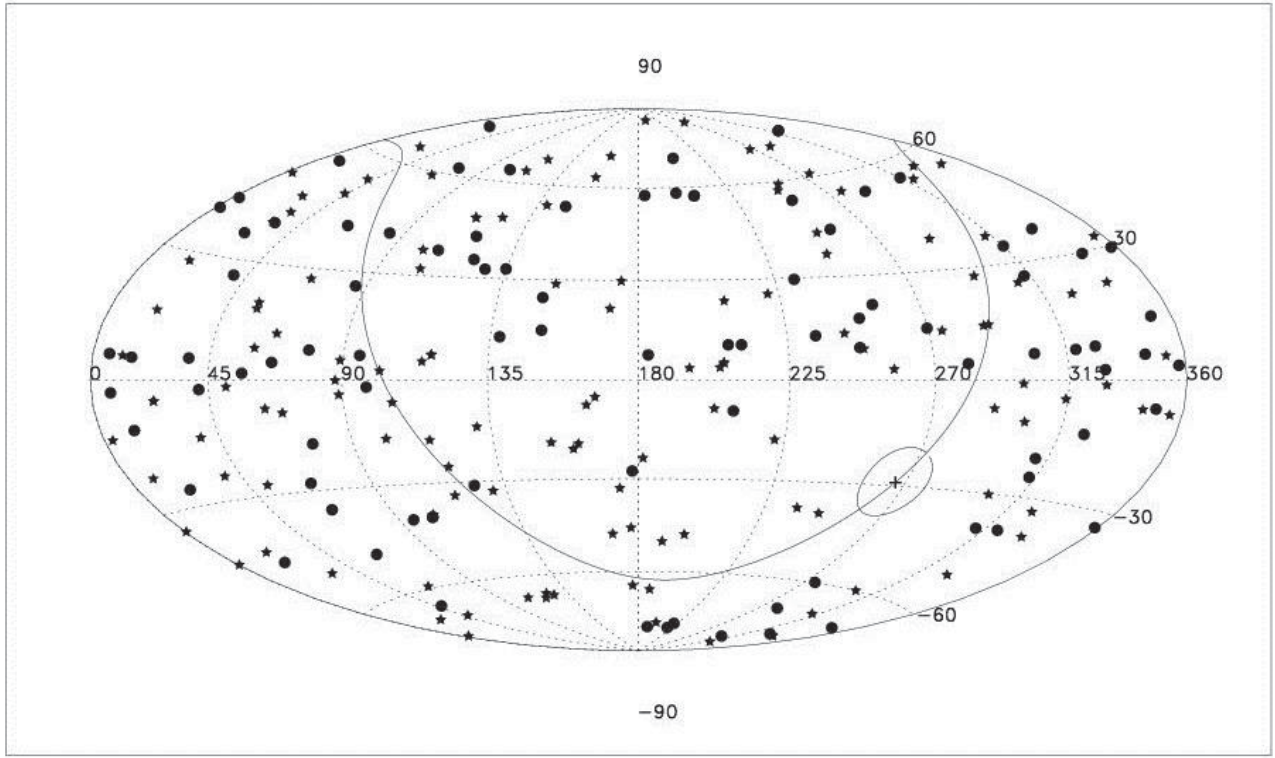

**Figure 1.** A Hammer-Aitoff projection of the local sample in equatorial coordinates centred on $12^h$ RA. The star symbols are within 20 pc and the circles are stars between 20 pc (i.e. LS20) and 25 pc (i.e. LS25). The Galactic plane is indicated by the solid curve. The location of the Galactic centre is indicated by a '+' symbol and the 'Galactic Bulge' is characterized by a circle of radius of 10 degrees.

It might be expected that the Galactic plane would exhibit a reduced areal density of local WDs due to field crowding and the difficulty of determining accurate proper motions. Indeed, Luyten (1979), the discoverer of many stars in the LS25 sample, often avoided fields of high stellar density at low Galactic latitudes. However, an examination of the areal density of local WDs as a function of Galactic latitude shows no overall convincing deficit of stars within |b| < 8.2° compared to zones of equal area located at higher latitudes. On the other hand, the largest void in Fig 1. is located in the direction of the Galactic center. This is similar to what can visually be estimated for the larger 40 pc sample of northern hemisphere WDs in Fig. 9 of LBL. We find the number of WDs in LS25 located in north of the Galactic plane to be 129 vs 103 located south of the Galactic plane.

*3.2 Distance Estimates*

Reliable stellar distance estimates are an essential aspect of this study. Fortunately most nearby WDs discussed here have good trigonometric parallaxes. Fully 81% of the stars in LS20 have measured parallaxes. However, this fraction drops to 68% of the LS25 sample. When possible, photometric distances (and uncertainties) were also calculated to supplement the trigonometric estimates. In general, the agreement between these two methods is excellent (HBG). Table 3 provides the adapted distances calculated from the trigonometric parallaxes (see Table 1) or photometric distances calculated from the adapted $T_{eff}$ and log g (Columns 2 and 3) photometry in Tables 2. In general trigonometric parallaxes are the preferred method of distance estimation. Column 6 in Table 4 gives the final adopted distance and Column 7 indicates the method based either on the trigonometric parallax (p) the spectroscopic distance (s) or in a few cases independent distance estimates. Notes on the adopted distances and other characteristics are given in Appendix A. These adopted distances are used in our construction of the sample. It should be noted that in Fig. 2 and in our discussion of the number of stars within various radial distances, as outlined in H08 distance uncertainties were used to calculate the probability that a star lies within a particular range of distances. This helps to reduce the sensitivity to the small number stars within some volume elements to simple enumerations based on specific distances. Membership of several DB WDs rests primarily on photometric estimates derived from high spectroscopic gravities with relatively high uncertainties. Taking these uncertainties into account helps to quantify the expected number of such stars in our 25 pc sample. It also allows the inclusion of stars that lay just beyond the sample distance limit of 25.2 pc that still have a significant probability of being within the LS25 sample.

Photometric distance estimates were computed for most DA, DAZ and DC stars using the spectroscopic $T_{eff}$ and log g values in Table 4, together with all existing multi-band photometry. The method employed is that described in HBG. An added improvement adopted here was to include the $T_{eff}$ and log g corrections discussed in Tremblay et al. (2013) which account for the well-known over-estimations of log g that associated with the standard Balmer line spectroscopy of DA WDs between 6000 K < T <

13,500 K. This correction has the net effect of lowering gravities and masses and thus increasing photometric distances estimates for these stars. It should be noted that the $T_{eff}$ and log g values adopted in Table 4 are the published values *but the distances and masses that result from them are derived from the Tremblay corrections*. In certain instances (see Appendix A) it is evident that the tabulated photometry refers to the composite light of an unresolved binary (e.g. RR Cae) or is contaminated by scattered light from a bright star (JHK magnitudes for 40 Eri B), or grossly incorrect. In these cases the magnitude uncertainties were set to large values so as not to bias the mean distances.

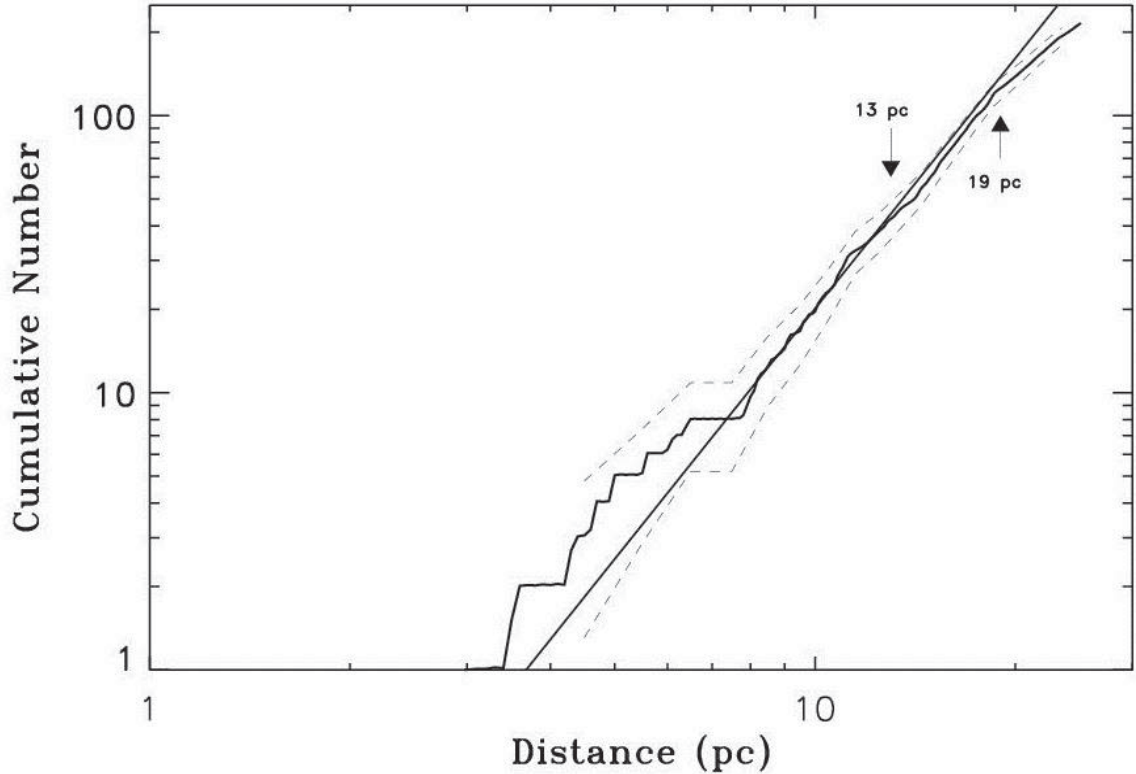

**Figure 2.** The observed cumulative distribution of nearby WDs as a function of distance. The irregular stepped curve is the observed cumulative distribution computed as described in the text, while the straight line is the expected distribution for an idealized uniform spatial distribution of stars. The dashed lines are the + 1σ and − 1σ uncertainties. Arrows indicate a distance of 13 pc where the sample is assumed to be virtually complete and a distance of 19 pc where the observed slope begins to deviate from +3, which would be expected from a complete sample.

*3.3 Physical Properties*

Table 3 gives the physical properties for each star in the LS25 sample. These include our adopted $T_{eff}$ and log g estimates along with the associated uncertainties and the dominant photospheric composition (Comp) noted as either hydrogen (H), helium (He) or mixed (H+He). Also listed are our derived WD parameters such as mass, bolometric magnitude ($M_{bol}$) and cooling age. The luminosities and ages are calculated from the Montreal WD photometric tables. The masses and luminosities in Table 3 are the basis the mass distribution in Fig. 3, and the luminosity distribution in Fig. 4 and the age distribution in Fig. 5.

A key aspect of the LS25 sample is its high level of completeness. Since each star in the sample is associated with a trigonometric parallax or a photometric distance estimate, it is possible to directly plot the cumulative number of stars as a function of distance. In Fig. 2 the log of the cumulative number of stars (log $\Sigma N$) is plotted vs the log of distance (log d in parsecs). Also shown is the log-log slope of +3 expected for a spatially uniform stellar distribution having a constant completeness. The observed cumulative distribution is not a simple enumeration of stars sorted into various distance bins but rather the sums of the individual probability distributions that define each star a fixed distance. For the relativity limited number of stars dealt with here, this approach reduces the sensitivity to stars included within or excluded from fixed distance boundaries when the true distances remain uncertain.

Both the space density of nearby WDs and the completeness of the sample can be obtained from this plot under the assumption that the sample of WDs is essentially complete out to a distance of 13 pc. The justification for this assumption is discussed below. An earlier version of this plot that extended to 20 pc was presented in LWD08. It is evident from Fig. 2 that sample completeness falls with increasing distance beyond 13 pc. The incompleteness fraction of the sample is a function of the difference between the observed curve in Fig. 3 and the line with a slope of +3. Under the assumption that the sample is complete to 13 pc, the incompleteness remains relatively constant at 10 % to 12 % between 13 pc and 19 pc before dropping to 30 % near 25 pc. Since there have been no significant recent additions to the 13 pc sample the space density of local WDs remains at $4.8 \pm 0.5 \times 10^{-3}$ pc$^{-3}$. The actual space density was determined by a least squares normalization of the +3 slope to the observed slope between 8 < d < 13 pc. However, a nearly identical density can be determined from the 43 WDs in the 13 pc volume. From the masses in Table 3, the corresponding local mass density is $3.1 \pm 0.3 \times 10^{-3}$ $M_\odot$ pc$^{-3}$.

Our conclusion that the local sample is complete out to a distance of 13 pc deserves some discussion, as it determines our estimate for the space density of local WDs. Given the low luminosity of most of the local WDs, it is impossible to know whether more undiscovered WDs may lurk within 13 pc or be undetected companions to nearby brighter stars. However, in spite of the explosive growth in the observed number of WDs over the last decade very few new WDs within 13 pc have been found. One recent discovery of a WD within 13 pc is WD 0416+593, a close Sirius-Like companion to ε Ret (Chauvin et al. (2006). This dearth of new WD discoveries continues even after comprehensive common proper motion surveys such as Lépine (2005) and the extensive northern hemisphere survey of WDs sensitive to very faint luminosities by LBL. LWD02 reported 46 stars closer than 13 pc, while in LWD08 there were 44 stars, the differences being attributable to stars migrating across the 13 pc boundary due to more precise distance estimates. In the current sample there are 43 such stars. It appears that not many truly nearby WDs remain to be discovered in this volume.

*3.3.1 Mass Distribution*

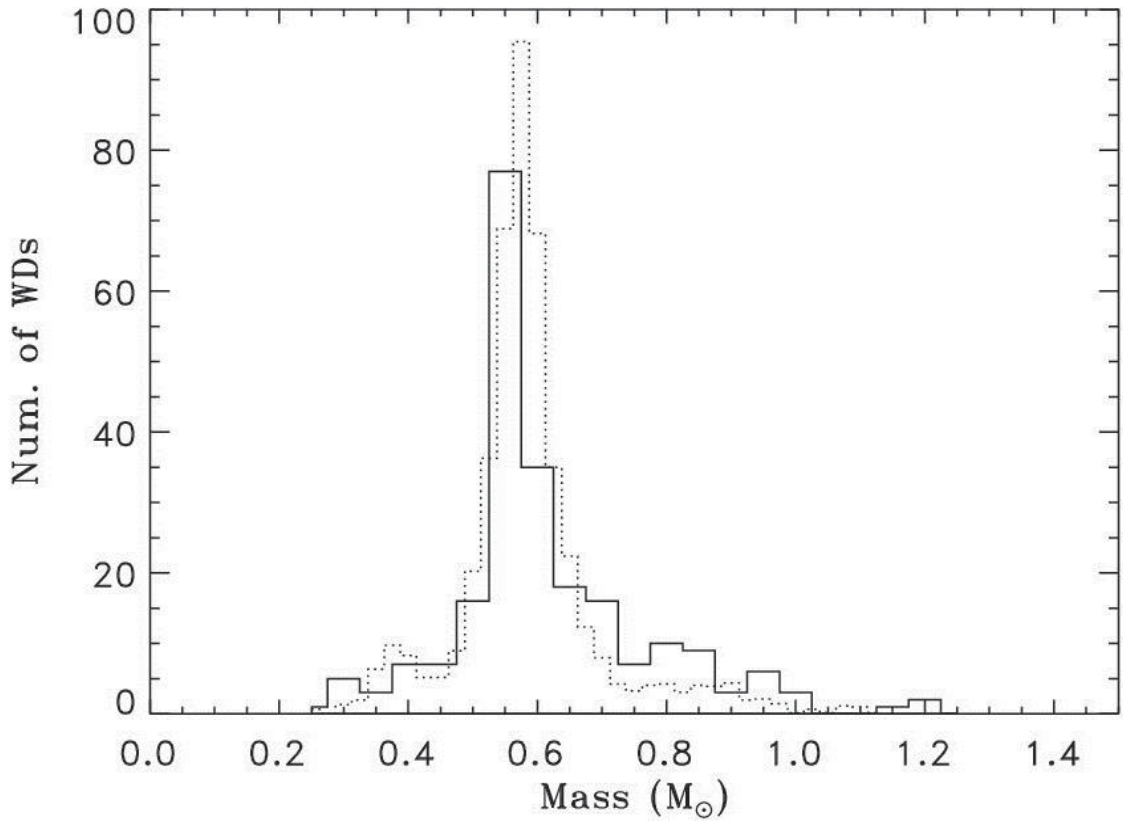

**Figure 3.** The LS25 sample mass distribution (solid histogram). The dotted histogram is from the Kleinman et al. (2013) DA mass distribution that has been normalized to the number of WDs in our LS25 sample. Our LS25 sample has been binned in 0.05 $M_\odot$ increments, while the larger Kleinman et al. sample is binned in 0.025 $M_\odot$ increments.

The observed LS25 mass distribution shown in Fig. 3 is compared to the appropriately normalized mass distribution from Kleinman et al. (2013) of the Sloan Digital Sky

Survey (SDSS) DR7 DA stars. The LS25 sample exhibits the familiar three-component mass distribution seen in larger spectroscopic studies of DA stars. A three-component Gaussian fit to the mass distribution yields a primary peak at 0.578 $M_\odot$, while a low mass enhancement associated with binary star evolution leading to He-core WDs is located near 0.35 $M_\odot$. There is also a significant high mass tail and a broad peak located near 0.82 $M_\odot$.

The 25 pc sample, being an amalgam of all spectral types, provides an inclusive representation of the WD mass distribution, rooted in a well-defined volume-limited sample. It exhibits the well-established characteristics of larger and fundamentally different magnitude-limited samples, such as that of Liebert et al. (2005). A mean mass, $<M> = 0.642$ $M_\odot$, is found for the full LS25 sample. This can be compared with two other recent volume-limited samples. Giammichele et al. (2012) find a mean mass of $<M> = 0.650$ $M_\odot$ for their 20 pc sample and Limoges et al. (2015) find $<M> = 0.699$ $M_\odot$ for their 40 pc sample, which the latter authors point out includes an increased number of high-mass WDs. Another point of comparison is large spectroscopic samples of DA stars. For example, Liebert et al., analyzing 298 DA stars determined $<M> = 0.629$ $M_\odot$, while Gianninas et al. (2011), analyzing over 1300 DA stars found $<M> = 0.638$ $M_\odot$. Limoges & Bergeron (2010), analyzing 136 DA stars in the Kiso sample, found $<M> = 0.606$ $M_\odot$. These latter examples were restricted to $T_{eff} > 13,000$ K, which effectively avoids the problem of cooler DA stars that are expected to yield larger than average spectrographic log g values, hence larger masses. The largest current sample, Kleinman et al. (2013), used 2217 DA WDs with an exclusive temperature range above 13,000 K to determine a mean WD mass of $<M> = 0.593 \pm 0.002$ $M_\odot$ and a peak mass of $M = 0.589$ $M_\odot$. As seen in Fig. 3 our peak mass (0.578 $M_\odot$) effectively coincides with the better defined Kleinman et al. peak mass but our mean mass is greater due to an enhanced fraction of higher mass stars.

*3.3.2 Luminosity Distribution*

In Fig. 4 the WD luminosity distribution (WDLF) for the LS25 sample is presented and compared with the Harris et al. (2006) WDLF. The latter was computed from reduced proper motions of 6000 WDs taken from the SDSS Data Release 3 survey and is binned in 0.5 magnitude increments. The local WDLF shown here uses $M_{bol}$ computed from the Montreal Photometric Tables[8] and normalized in the following fashion. The stars were placed into full magnitude bins and normalized by the total volume of the 25.2 pc sphere of the LS25 sample. The number of stars in each bin is corrected for distance-dependent incompleteness by assigning each star a number representing the reciprocal of the completeness at that distance (see Section 3.2). Thus, for stars with distances less than 13 pc this completeness number is 1.0 while for stars between 13 pc and 19 pc where the incompleteness averages 0.82 the number is 1.0/0.82 = 1.22 and so on out to 25.2 pc where the incompleteness approaches 0.7. The maximum observed luminosity in the LS25 sample is $M_{bol}$ = 7.55 (WD 1620-391) while the minimum luminosities are $M_{bol}$ = 17.6 and 17.69 (WD 1401+457 and WD 2058+342, respectively).

---

[8] http://www.astro.umontreal.ca/~bergeron/CoolingModels/

The Harris et al. WDLF shown in Fig. 4 covers a solid angle of approximately one eighth of the sky. It is corrected for incompleteness and for the effect of interstellar reddening at distances beyond 100 pc. Since it extends to significant distances it also includes a WD Galactic scale height, which for the data in Fig. 4 was assumed to be 250 pc however, the analysis of Harris et al. favours WD scale heights above 300 pc. Our 25 pc WDLF is in reasonable agreement with the Harris et al. WDLF, considering the individual error bars of the points in our LS25 sample. Although a much smaller survey, the LS25 sample contains no dependence on WD Galactic scale height or corrections for interstellar reddening and only modest distance dependent corrections for sample completeness. It also differs from the Harris et al. WDLF by including WDs in Sirius-Like and double degenerate systems. Finally, our WDLF also is also seen to be very similar to the respective 20 pc and 40 pc WDLFs of Giammichele et al. (2012) and Limoges et al. (2015).

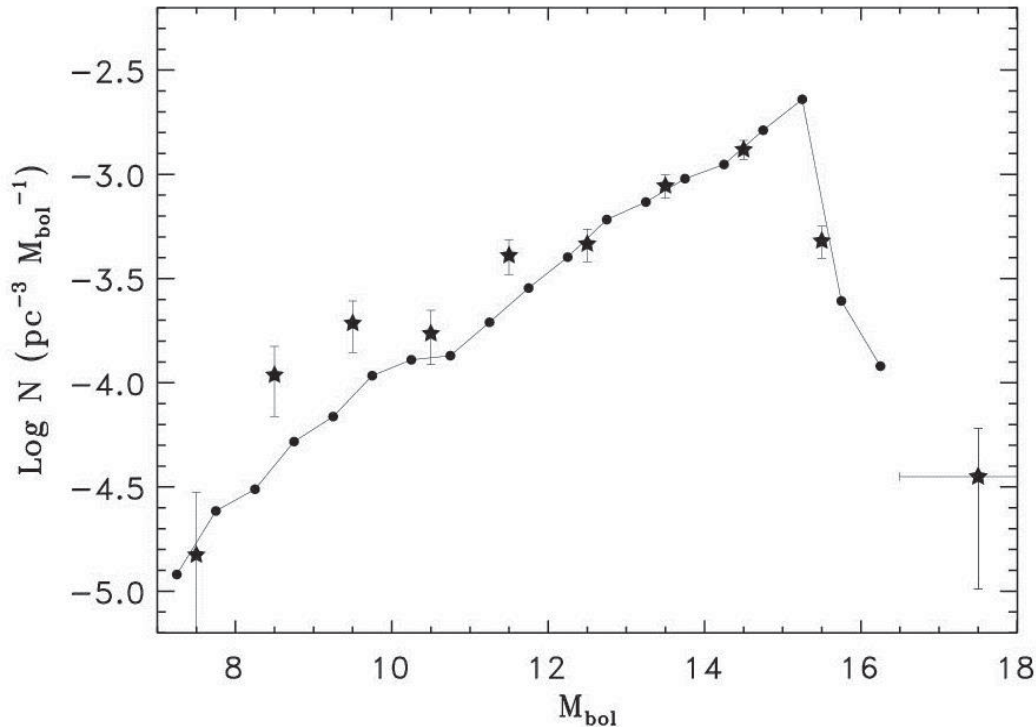

**Figure 4.** The LS25 sample luminosity function (large stars with error bars based on √N) compared with the Harris et al. (2006) luminosity function (filled circles). The single low luminosity point at $M_{bol}$ = 17.5 (with horizontal and vertical error bars) represents the two lowest luminosity stars in our sample.

In our sample there are no stars in the 16.0 to 17.0 $M_{bol}$ bin, but two WDs with poorly determined luminosities are in the 17.0 to 18.0 $M_{bol}$ bin (see above). We have chosen to represent theses with a single point at $M_{bol} = 17.5$ accompanied by a large horizontal error bar. For reference, each star in our bins represents an increment of N by $1.76 \times 10^{-5}$ pc$^{-3}$ $M_{bol}^{-1}$.

*3.3.3 Ages and Birth Rates*

For each star in the LS25 sample it is possible to estimate a cooling age. For the DA, DZ and DC stars cooling ages were interpolated from the Montreal Photometric Tables (see footnote 8). It should be noted that for these stars our ages included corrections for convection suggested by Tremblay et al. (2013) appropriate for stars 6000 K < $T_{eff}$ < 13,500 K. For non-DA stars, such as DZ, DQ, and others, the ages given by GBD or LBL were used. In Fig. 5 the frequency distribution of cooling ages has been scaled to represent an apparent 'WD birth rate' as a function of cooling age. The bin widths in this figure are 0.25 Gyr and the volume is that of the LS25 sample, including a correction for sample completeness. In this representation of the LS25 sample cooling ages the origin is the present and WD birth rates extend back more than 8 Gyrs.

There have been numerous prior estimates of the WD birthrate, using both indirect determinations from planetary nebulae (PN) birth rates and more direct determinations from large WD spectroscopic samples. For example, Liebert et al. (2005) estimate a current birth rate of $6 - 7 \times 10^{-13}$ stars pc$^{-3}$ yrs$^{-1}$ for the PG sample of DA stars and ~ $10 \times 10^{-13}$ stars pc$^{-3}$ yrs$^{-1}$ for all WDs. Vennes et al. (1997) found a birth rate of $7 - 10 \times 10^{-13}$ stars pc$^{-3}$ yrs$^{-1}$ for their EUV selected sample of DA WDs. Other estimates using various samples and techniques are to be found in Verbeek et al. (2013), (see their Table 4, for a list of prior determinations). It should be noted that almost all these estimates refer to the present epoch (< 1 Gyr), with little or no information about the birth rates in earlier epochs. Moreover, most of the prior spectroscopic samples are restricted to DA stars with $T_{eff}$ > 13,000 K and/or absolute magnitudes brighter than $M_{bol}$ < 12.5. The LS25 sample includes WDs of all spectral types and temperatures.

We estimate from the LS25 sample that the present birth rate during the past 500 million years is ~ $14 \times 10^{-13}$ stars pc$^{-3}$ yrs$^{-1}$. Taking the solid line in Fig. 5 at face value would imply that the present WD birth rate has increased by two or threefold over 8 Gyr. The small number of stars beyond 6 Gyrs limits confidence in these birth rates. It is of course possible to link these WD birth rates to the prior history of star formation in the Galactic disk. Such a study would require linking individual WD masses (which are available) to a robust model of the Initial Mass – Final Mass Relation (IMFMR). It would also require including only WDs that are the product of single star evolution.

Thus, no attempt is made here to derive the corresponding stellar formation rates. However, Tremblay et al. (2014) have considered this issue using a different approach.

Although the cooling ages in Fig. 5 should represent a reasonably accurate time scale, the birthrates must be modified by the gravitational "scattering" of older stars into a phase space characterized by larger velocity dispersions and thus larger scale heights. In effect, for the volume-limited sample discussed here the overall apparent space density for older stars will decrease with time as the corresponding space motion dispersions increase. This effect was quantified, to first order, by dividing the LS25 sample into an "older" component having $T_{eff} \leq 8000$ K (age $\geq 1.37$ Gyr.) and a "younger" sample with $T_{eff} \geq 8000$ K (age $\leq 1.37$ Gyr.). For each sample space motions were computed in the manner described by Sion et al. (2014). As expected, the velocity dispersions were found to be significantly larger for the "older" sample than for the "younger". In particular, the local vertical velocity dispersions for the 'older' sample is $W = 35$ km s$^{-1}$ vs $W = 24$ km s$^{-1}$ for the 'younger' sample. A uniform scaling factor for the older sample was computed under the assumption the volume dilution between the two samples is proportional to the ratio of their galactic scale heights and that the scale heights are in turn proportional to the squares of the respective vertical velocity dispersions in each sample. This leads to a dilution factor of 2.26 between new and old sample. This corresponding effective increase in the birthrate is shown in Fig. 5 as star symbols. As can be seen this tends to diminish the age-dependent gradient in the WD birthrate. With a larger volume-limited (i.e. > 25pc) sample of stars it should be possible to better define the local WD birth rate function.

**4.0 ANALYSIS**

**TABLE 4**
**WD Populations – Binary and Magnetic**

| Sample | R < 20pc | 20 pc < R <25 | Total |
|---|---|---|---|
| **Volume** | **33.5x10$^3$pc$^3$** | **33.5x10$^3$pc$^3$** | **67x10$^3$pc$^3$** |
| **All** | **136** | **96** | **232** |
| **Single** | **95** | **79** | **173** |
| **Binary** | **15** | **9** | **24** |
| **Sirius-Like** | **11** | **1** | **12** |
| **Double Degenerate** | **14** | **10** | **24** |
| **Magnetic (DP/DH)** | **16** | **5** | **21** |

The LS25 sample can be simply divided into two independent samples of equal volume; one a sphere of radius 20 pc and the second a spherical shell with inner and outer radii of 20 pc and 25.2 pc. This allows a direct comparison of the relative populations in each sample, that helps point to which category of WDs might be missing from the current overall sample. As expected, cooler lower low luminosity stars are more numerous in the 20 pc subsample compared to the 20 pc to 25 pc subsample. For example, for $T_{eff}$ < 6000K there are twice as many such stars in the interior sample than in exterior sample, 49 vs 25 stars, respectively. Table 4 lists the population subtypes found in our current LS25 sample.

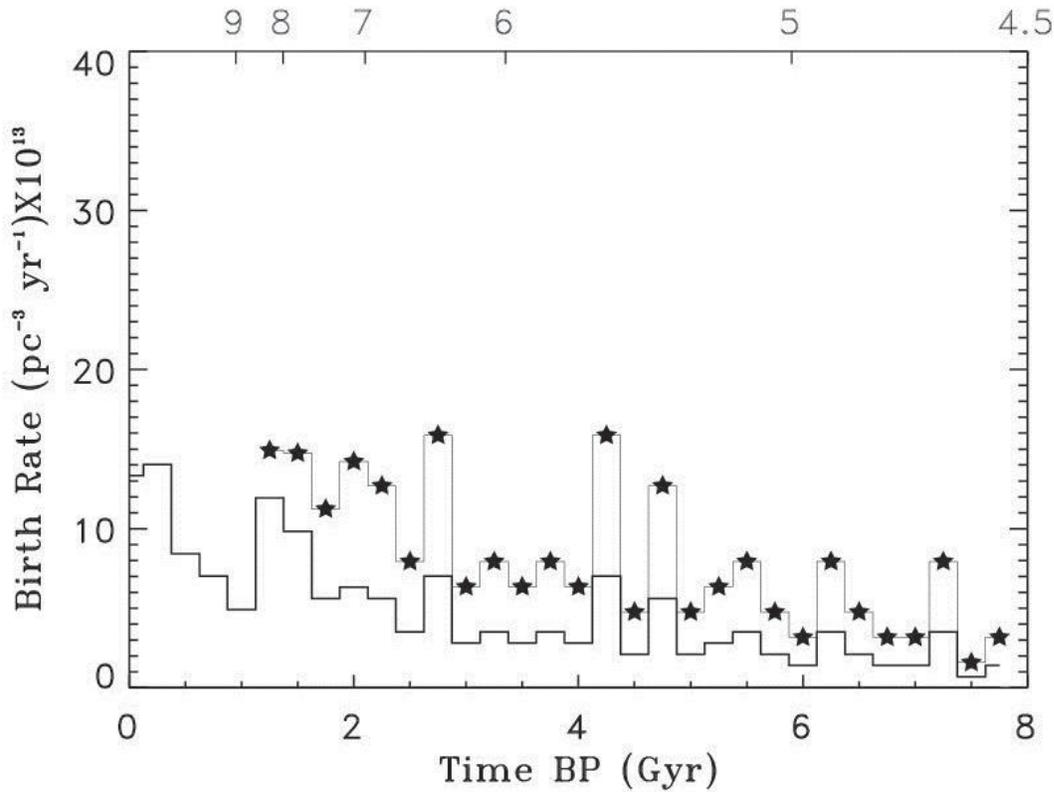

**Figure 5.** A plot of the local WD birth rate as a function of time before present (TBP) in Gyrs, Data has been binned in 0.25 Gyr increments. The labels above the histogram indicate the mean effective temperatures as a function of age. The histogram marked with stars indicates a first order correction based on higher dispersion velocities for older stars as outlined in the text.

*4.1 Binary and Multiple Systems*

Of the 232 individual white WDs in our LS25 sample, 60 are in systems containing more than one star, including wide binaries. In order to avoid the ambiguity in counting double degenerate (dd) systems our 'binary fraction' is defined as unity minus the fraction of single WDs. Using this definition, a fraction of only 26 % of the stars in our local sample are members of binary or multiple systems[9]. Three basic categories of binary or multiple star systems in the LS25 sample are: double degenerate systems, WDs with M-star companions and Sirius-Like systems in which the WD is a companion to more luminous K or earlier type star. Table 4 lists 24 WD + M stars, 12 Sirius-Like systems and 24 WDs in resolved and unresolved double degenerate systems. Compared to the 20 pc sample, many Sirius-Like systems, a considerable fraction of binary systems and 'dd' systems between 20 pc and 25 pc have apparently gone undetected.

Our finding that 76% of the local WD population consists of single stars is a robust statistic that is unlikely to change significantly with the addition of more stars to the local sample or the advent of even larger samples. This contrasts with many well-accepted studies that suggest the majority of stars are members binary or multiple star systems. For example, (Mason et al. 2009) found that ~ 25 % of O and B stars are single, while De Rosa et al. (2013) in studying multiplicity among A-type stars within 75 pc, found 56 % are single stars. Considering later type stars, Raghavan et al. (2010) found that for F6-K3 stars within 25 pc, $52 \pm 2$ % are single. Only M-type stars (which have not formed WDs over the age of the Galaxy) have higher rated of singularity. Reid & Gizis (1997) in a study of M-type stars in the Hyades found that 88.7% $\pm$ 4.6 % are single. Thus the discrepancy posed by our study of the LS25 sample that at most 26 % of local degenerates can be explicitly regarded as being members of multiple systems, is in contrast to theoretical and empirical studies of local main sequence stellar populations that have found much larger fractions. The significance of this seeming paradox is somewhat lessened by the discussion in Section 4 which clearly implies that there remain substantial numbers of undiscovered local WDs in Sirius-Like and dd systems. Given that WDs represent the descendents of earlier generations of main sequence stars, one would expect, in the absence of significant evolutionary effects, such as close binary evolution and mass exchange, that the present WD population should reflect a higher fraction of multiple stars. However, before it can be concluded that the present local WD population has abnormally low binarity or multiplicity careful consideration needs to be given to observational biases that result in binaries being overlooked or evolutionary effects that lead to the disappearance or nondetection of companions.

Is it plausible that the deficit of WD binaries provides observational evidence that former binary companions have been 'lost' in the processes that lead to the formation of WDs? Such binaries can be dissociated by outright impulsive mass loss subjected to dissolution by the Galactic potential, and/or encounters with stars and giant molecular clouds. This has been investigated by Johnston, Oswalt, & Valls-Gabaud (2012) who

---

[9] Multiple stars systems: WD 0322-019, WD 0413-077, WD 0727+482, WD 2058+342

Systems with sub-stellar components: WD 0210-508, WD 0416-593, WD1620-390

found that such critical orbital distances are on the order of $10^4$ AU. They also found that the Luyten proper motion sample matches the characteristics of simulated populations of evolved binary systems.

*4.2 Sirius-Like Systems*

Holberg et al. (2013) discussed the question of Sirius-Like Systems (SLSs), defined as binary or multiple systems containing a WD with a more luminous companions of spectral type K or earlier. They found 98 known systems of this type ranging over distances from 2.6 (Sirius B) to 300 pc. Notably, they found 11 SLSs within the volume of the LS20 sample, but none were found in the volume between 20 pc and 25 pc. Indeed, only one SLS was found between the 20 pc sphere, and the next four consecutive spheres of equal volume. This study concluded that many nearby SLSs have been overlooked. From the LS20 sample Holberg et al. estimated a SLS space density of $3.4 \times 10^{-4}$ pc$^{-3}$ and that such systems occur with a frequency of 0.6 % to 1.2 % among nearby main sequence stars of spectral type K to B. The presence of WD 2307-691, a previously unknown companion to the K3V star HD 218572 which has a *Hipparcos* distance of $20.94 \pm 0.38$ pc is noted in this paper. The existence of this system was found by Brian Skiff (2013, private communication) and represents the first unambiguous Sirius-Like system between 20 and 25 pc. A second SLS, the G5V star HD 114174 which lies at $26.14 \pm 0.37$ pc has been discussed by Crepp et al. (2013). Clearly, the implication is that others remain to be discovered.

*4.3 Double Degenerate Systems*

There are 12 double degenerate systems (i.e. 24 WD components) in the LS25 sample; seven are within 20 pc and the remaining three are beyond 20 pc. Of the total, five are unresolved systems and seven are resolved. For the resolved systems the photometric distances for each component are consistent with the physical proximity (and where the data exist) also consistent with the parallax distance. Two of the unresolved systems have been observed to display short period large amplitude radial velocity variations: WD 0135-052 (1.556 days, Bergeron et al. 1989) and WD 0326-273 (1.88 days, Nelemans et al. 2005). From the 20 pc portion of our LS25 sample, the space density of double degenerate systems is at least $6.2 \pm 0.5 \times 10^{-4}$ pc$^{-3}$.

*4.4 Magnetic Degenerates*

Magnetic WDs are an intriguing component of the local WD population, comprising about 9% of the full LS25 sample. The presence of magnetic fields in WDs is most frequently manifested by either Zeeman splitting (DH stars) or polarization (DP stars). In the 20 pc portion of our LS25 sample 16 out of 136 WDs (12%) are magnetic and five out of 93 are in the 20 pc to 25 pc volumetric shell. There are good reasons to regard these fractions as firm lower limits.  First, stars cooler than 5000 K effectively have no Hα line, ruling out detection by Zeeman splitting.  Second, many of the cooler and more recently discovered LS25 WDs have not been inspected for polarization (see Landstreet et al. 2012).  This is consistent with Table 4 which shows that the majority of the known magnetic WDs are in the 20 pc portion of the full LS25 sample.

A plausible scenario that can lead to 'lost' binaries is the merger of a close binary during the common envelope phase, resulting in a more massive and possibly magnetic WD.  A decade ago Liebert et al. (2005) discovered that in detached WD + dM systems, the WDs are exclusively non-magnetic, while known magnetic WDs (MWDs) have no detected main sequence companions.  These findings have been dramatically confirmed with even larger WD samples (Liebert, et al. 2015).   Several theoretical studies (see for example Tout et al. - 2008, and Nordhous et al. - 2011) have suggested that isolated magnetic WDs are the result of binary mergers which result in a single MWD.  In this process a common envelope dynamo produces a magnetic field that is retained by the WD.  Adopting the view that all MWDs in our sample were once members of binary systems, has the effect of raising the fraction of present or former binaries to over 33%. Using this idea, the 9% of the LS25 population that are magnetic could be regarded as having once been close binary systems.  The mean mass of the 21 magnetic WDs in our sample is $M \sim 0.70$ $M_\odot$, significantly larger than the mean mass for the local sample as a whole (see section 3.3.1).  One would expect that not only would the WD emerge from the common envelope phase with a magnetic field but it would also retain a fraction of the mass of the companion.  This strengthens somewhat the notion that they may have arisen from mergers involving high mass loss.

*4.5 DB Stars*

The LS20 sample of H08 contained no spectroscopically-identified DB WDs.  Bergeron et al. (2011), however, conducted a comprehensive analysis of 108 DB stars including spectroscopic distance estimates. They found four DB stars nominally within 25 pc: WD 1542-275 (22 pc), WD 2058+392 (24 pc), WD 2147+280 (20 pc), and WD 2316-173 (11 pc). All have surface gravities ($\log g > 8.8$).  Of these, WD 2147+280 has a trigonometric parallax (van Altena et al. 1994) suggesting a distance of $35.7 \pm 0.4$ pc, which as the authors point out, is difficult to reconcile with the estimated distance unless the gravity is reduced to 8.2 in which case it becomes consistent with the larger distance.  All four stars appear to be massive ($M > 1.1$ $M_\odot$) and have similar temperatures ($T_{eff} \sim 12,000$ K). They also possess similar gravities ($\log g \sim 9.1$) and have relatively large gravity uncertainties ($\Delta \log g > 0.15$), implying larger than average distance uncertainties.  As pointed out by Bergeron et al. it is difficult to determine accurate gravities when the He I lines become weak.  Moreover, in their plot of mass vs $T_{eff}$ (their Fig. 21) there is an obvious increase in the number of high mass stars with $T_{eff} < 18,000$ K. This leads to the suspicion that the masses, and gravities of DB WDs may be biased towards larger values and hence to smaller distance estimates. An alternative to the spectroscopic gravities used by Bergeron et al. (2011) is to employ the distribution of log g values for DB stars from Kleinman et al. (2013) to calculate a distribution of distances. This procedure produces the following distance estimates; WD 1542-275 (53.8 pc ± 11.0), WD 2058+392 (60.7 ± 12.7 pc), and WD 2316-173 (29.4 ± 6.0 pc). On balance we conclude that none of these Bergeron et al. DBs belongs to the 25 pc sample. Overall, Bergeron et al. found a DB space density of 5.15 x $10^{-5}$ $pc^{-3}$ which gives an expected number of roughly 3 - 4 DB WDs in our 25 pc sample, which currently contains no DBs.

There are two other DB stars worth mentioning with published distances placing them within 25 pc: WD 0503+147 (22 pc) and WD 0615-591 (23.8 pc). WD 0503+147 was originally included in Limoges & Bergeron (2010), but its revised distance in Bergeron et al. (2011) is now 31 pc. WD 0615-591 has a parallax by van Altena et al. (1994) placing it within 25 pc; however, Gould & Chanamé (2004) identified a *Hipparcos* common proper motion companion which suggested its parallax distance is actually 36.4 pc. Neither of these stars has been included in our LS25 sample.

## 5.0 'MISSING' STARS

From our estimate of the observed space density of the local sample together with the expected number of WDs for a uniform density, the number of WDs missing from the current sample can be estimated. This missing fraction turns out to be approximately 100 WDs, or an overall fraction of ~30%. The question we now address is where and how might these missing members might be found? As mentioned in Sec. 3.1, there exists a significant excess of northern hemisphere stars in the current sample. LS25. Based on this disparity, improved southern hemisphere proper motion surveys might be expected to add an extra ~ 30 WDs. Another obvious source of missing WDs is undiscovered Sirius-Like systems. As discussed in Section 4.2 it is reasonable that perhaps as many as 10 such systems may exist between 20 pc and 25.2 pc. The recent discovery of WD 2307-691 at 20.9 pc discussed above suggest that more resolved Sirius-Like discoveries remain to be discovered. However, most remaining Sirius-Like systems are likely to be difficult or impossible to resolve (see Holberg et al. 2013 for a discussion how such systems are likely to be detected or inferred). Another possible source of additional WDs in the local sample is double degenerate systems. Here close inspection of existing local WDs may reveal radial velocity variations or significant mis-matches between trigonometric distances and photometric distances. However, our study of the current local sample shows no such cases that were not already known. Additional Sirius-Like or double degenerate systems would indeed increase the overall binary fraction but not close the previously mentioned gap.

There exist several binary systems in the literature that are within 25 pc that may contain WDs. As mentioned in Holberg et al. (2013) the bright star Regulus was reported to be a single line binary by Gies et al. (2008). It is a strong possibility that the unseen companion is a WD. There is also a DC WD + M5.0V, binary with an estimated distance of 24.6 pc, discovered by Jao et al. (2014). The cool WD (4800 K ± 200 K) was detected astrometrically by the reflex motion of the M dwarf and photometrically using *HST* in the 1820 Å to 2700 Å band. Astrometry yields an estimated orbital period P = 15.17 ± 3.5 yrs and an apparent separation of 0.2082". The estimated mass of this object, WD 1840+042 is uncertain, but its magnitude is estimated to be V ~ 17.4. It is not included in this sample due to its uncertain distance.

Fig. 2 indicates that most of the new local WDs will found be at distances between 20 pc and 25 pc. It is likely that they will have uncharacteristically low proper motions or have small radii and be more massive than average (see Limoges et al. 2015) as well as cooler WDs (see Section 4.0). Ultimately the *Gaia Mission,* through direct parallax determinations, should find nearly all of the WDs within 25 pc and well beyond. Many of these will be between of $15 < M_v < 16$ at and beyond the peak of the WD luminosity function. Among those that elude *Gaia* will be some very close Sirius-Like systems, unless the companions can be deduced from astrometric variations.

## 6.0 CONCLUSIONS

Extending the local WD population from 20 pc to 25 pc not only doubles the volume of space, but it also nearly doubles the sample size. The resulting larger sample of 232 stars helps strengthen a number of conclusions about the LS20 sample. For example, the space density remains at $4.8 \pm 0.5 \times 10^{-3}$ pc$^{-3}$, a result that is likely to change only if there are substantial undiscovered WDs within 13 pc. Another result established by our study of the LS25 sample here is a low (~ 25 %) fraction of binary and multiple systems. The implication of this is that the post-main sequence evolution of binary systems results in a significant loss of companions, leaving a single WD. Theoretical studies suggest that the origin of magnetic WDs is associated with common envelope evolution in which the WDs acquire a magnetic field from a dynamo and the low mass companion merges with the WD. If it can be confirmed that most local magnetic WDs were once close binaries then the shortfall of WDs found in binaries would be explained. Our local WD mass distribution presented here has a mean mass of <M> = 0.642 M$_\odot$ and a peak mass of 0.578 M$_\odot$. Our local WD luminosity distribution, valid from $7.0 < M_{bol} < 16.0$ is in good agreement with the Harris et al. (2006) luminosity distribution.

## Appendix A

**WD0000+345**
Reimers et al. (1996) find Zeeman splitting and estimate a field of 43-118 MG for this star.
**WD0009+501**
This is a magnetic WD with $B_p$ ~ 0.2 MG and reported photometric variability (see Brinkworth et al. 2013 for discussion).
**WD0011-134**
This is a magnetic WD with $B_p$ ~ 16.7 MG and reported photometric variability (see Brinkworth et al. 2013 for discussion).

**WD0101+048**
This is an unresolved double degenerate system (Zuckerman et al. 2003, Koester et al. 2009). The trigonometric parallax gives a distance of 21.32 ± 1.73 pc, while the UBVRI+JHK and *ugriz* photometry gives an internally consistent distance of 24.07 ± 0.64 pc. The slight difference between the parallax and photometric distances indicates one component is photometrically dominant.

**WD0108+277**
There exist various determinations of the parameters for this star (NLTT 3215, LP 294-61). Kwaka & Vennes (2006) find it to be a cool DAZ star, and use the weak Hα and Hβ lines to deduce a relatively high mass with $T_{eff}$ = 5270 ± 250 K and log g = 8.36 ± 0.60 which yields a spectrometric distance of 14.88 ± 2.00 pc. Farihi (2009) notes the presence of a nearby (2") fainter 'background star', which may have influenced the differences between the 2MASS JHK and IRTF JHK values. GBD) obtain a temperature of 6828 ± 190 K, but assign a canonical log g = 8, yielding a distance estimate of 28.0 ± 1.5 pc. The corresponding distance estimate obtained here is 34.2 ± 5.2 pc. Giammichele et al. suspect the star is most likely an unresolved DA + DC system. The true distance and nature of this star must await a trigonometric parallax.

**WD0121-429**
This is a likely an unresolved double degenerate system in which one component is magnetic (Subasavage et al. 2007). The photometrically dominant component is a DAH star showing Zeeman splitting. However, as pointed out by Subasavage et al., the low mass of 0.43 ± 0.03 $M_\odot$ (0.41 ± 0.02 $M_\odot$, here) indicates that the star is the product of common envelope evolution. Subasavage et al. note the presence of a diluted Hα core and attribute this to a cooler DC companion. We assign a plausible mass of 0.6 $M_\odot$ to the unseen companion.

**WD0123+732:** Oswalt & Strunk, (1994) list this as a DAB/dM5 pair. Our adopted distance is 25.1 ± 3.7 pc from Limoges et al. (2015).

**WD0135-052**
This was discovered as an unresolved double degenerate DA + DA system (Saffer, Liebert, & Olszewski 1988) with an orbital period of 1.556 days. Bergeron et al. (1989) find $T_{eff}$ = 7470 ± 500 K, log g = 7.80 ± 0.10, M = 0.47 ± 0.05 $M_\odot$ and $T_{eff}$ = 6920 ± 500 K, log g = 7.89 ± 0.10, M = 0.52 ± 0.05 $M_\odot$, respectively, for the two components. In this paper we have independently estimated the luminosities and ages as $M_{bol}$ = 12.84 and $M_{bol}$ = 13.30 and 1.07 Gyr and 1.42 Gyr, respectively using these parameters. The parallax gives a distance of 12.35 ± 0.43 pc while the composite UBVRI+JHK and *ugriz* photometry gives an internally consistent distance of 8.91 ± 0.13 pc. It is possible to estimate the individual photometric contributions of each star by using the Bergeron et al. $T_{eff}$ and log g estimates along with the constraints imposed by the trigonometric parallax distance and the observed magnitudes. This yields V magnitudes of 13.47 and 13.89 for the components WD0135-052A and WD0135-052B, respectively.

**WD0148+641**
This is the close DA 5.6/ M2V system (G244-37/36). The 2MASS JHK photometry is contaminated by nearby M dwarf companion.

**WD0210-508**
This is the close binary Sirius-Like system GJ 86 containing a WD and an exoplanet (see Farihi et al. 2013).

**WD0322-019**
This is an unresolved double degenerate system. It was suggested as being a possible low mass binary system by Leggett Ruiz & Bergeron (1998). Zuckerman et al. (2003) detected Hα radial velocity variations and double Ca II lines. Our internally consistent UBVRI+JHK photometry distance matches the parallax distance to within mutual uncertainties, indicating the DAZ is photometrically dominant.

**WD0326-273**
This is an unresolved double degenerate (DA+ DC) system as well as a triple system containing a distant M3.5 star (L 587-77B). Zuckerman et al. (2003) detected Hβ radial velocity variations in the DA. Nelemans et al. (2005) investigated the system as a single line spectroscopic binary, determining a gravitational redshift with respect to L 587-77B, a spectroscopic $T_{eff}$ and log g determination, and obtaining an orbital period of 1.88 days. A mass of 0.51 ± 0.05 $M_\odot$ was estimated for the DA star and a mass estimate of > 0.59 $M_\odot$ for the secondary. The unseen component is estimated to be a cool DC star, and based on the observed VRI+JHK excess. The trigonometric distance is 17.36 ± 4.1 pc and our VRI+JHK photometric distance is 21.04 ± 2.2 pc, thus indicating that there is indeed extra light in the dd system.

**WD0413-077**
This is the well known Sirius-Like system containing 40 Eri B. The other two components are the distant KV primary 40 Eri A and a closer M star (component C). The 2MASS photometry is contaminated by the primary.

**WD0416-593**
This is the Sirius-Like system ε Ret (HD 27442 B). Our results are taken from the ground based study of Farihi et al. (2011).

**WD0419-487**
This is the well known eclipsing variable RR Cae involving a DA and M4 star in a 0.30 d orbit. The parameters used here, including the $T_{eff}$ and log g, are taken from the comprehensive analysis Maxted et al. (2007) who obtain a mass of 044 ± 0.023 $M_\odot$. The JHK photometry is dominated by the M star.

**WD0423+044**

This is a cool DA with a parallax distance of 20.72 ± 1.66 pc from Gatewood & Coban (2009). It has no spectroscopic gravity and assuming log g = 8 yield a photometric distance of 26.07 pc, indicating that it likely has a higher gravity and mass.

**WD0426+588**

This is the DC star Stein 251B. Dietrich et al. (2012) give a separation of 9.201" and a Position angle of 63.7° for the M star binary component. The WD appears to have an IR (JHK) excess.

**WD0431-360**

Subasavage et al. (2009) estimate a distance of 25.2 ± 4.1 pc and a $T_{eff}$ = 5153±121 K. Our photometric distance is 25.67± 0.62 pc.

**WD0454+620**

This is a DA + M from Limoges, Lepine & Bergeron (2013) who estimate a photometric distance of 24.9 ± 0.9 pc. The JHK photometry is dominated by the M star.

**WD0503-174**

Kwaka et al. (2007) list a $B_p$ of 7.3 ± 0.2 MG for this star.

**WD0532+414**

This is an unresolved double degenerate system, where Zuckerman et al. (2003) have noted the presence of double $H_\beta$ cores. In the absence of follow on studies we assume the two components are similar DA stars, giving a photometric distance of 19.8 pc.

**WD0548-001**

This is a magnetic WD with $B_p$ ~ 10 -20 MG and a photometric period of 4.117 hr. (see Brinkworth et al. 2013 for discussion

**WD0553+053**

This is a magnetic WD with $B_p$ = 20 MG and a photometric period of 26.8 min. (see Brinkworth et al. 2013 for discussion).

**WD0628-020**

The 2MASS photometry belongs to the M star companion LP 600-43.

**WD0642-166**

Sirius-Like system; Sirius A and B.

**WD0651-398A/B**

This is a widely separated dd system (Subasavage et al. 2008). These authors find a spectroscopic $T_{eff}$ = 7214 ± 135 K and log $g$ = 7.68 ± 0.19 and an estimated distance of 25.1± 4.3 pc for component A. The photometric results for component B are $T_{eff}$ = 6450 ± 220 K and a distance of 26.9 ± 4.3 pc. We adopt the spectroscopic distance for both components.

**WD0659-063**

The existing trigonometric parallax for this star (van Altena et al. 1994) is π = 81 ± 24.2 mas (12.2 ± 3.7 pc). However, our photometric estimate is 17 ± 0.5 pc, which is more in line with a recent parallax measurement which indicates a distance of ~20 pc (Subasavage 2015, private communication). We use our photometric distance.

**WD0727+482**

This is a well observed close double degenerate system G107-70AB. Recently Nelan, Bond, & Schaefer (2015) have published results from the *HST* Fine Guidance Sensor (FGS) that provides an accurate astrometric orbit for this system as well as an improved parallax. The results used here are from Nelan et al. include π = 87.41 ± 0.48 mas, Orbital Period = 18.84 (.02) yrs, orbital semi-major axis = 663.62 ± 0.79 mas and the astrometric component masses of $M_A$ = 0.634 ± 0.01 $M_\odot$ and $M_B$ = 0.599 ± 0.01 $M_\odot$. G107-70AB is also part of a quadruple system including G107-69, itself a very close astrometric binary with an sDM primary with a 0.94 yr orbital period (Harrington, Christy & Strand, 1981). G107-69 has a separation of 103.2" and a position angle of 206.6° with respect to G107-70AB.

**WD0728+642**

This is magnetic WD with $B_p$ ~ 0.1 – 0.4 MG (see Brinkworth et al. 2013 for discussion).

**WD 0736+053**

This is the well known DQZ WD Procyon B. Recently, Bond et al. (2015) have presented a new fundamental re-analysis of the Sirius-Like system Procyon A/B and its evolution based primarily on *HST* astrometry. From Bond et al. we have adopted a mass of 0.592 ± 0.006 $M_\odot$, a radius of 0.01234 ± 0.0032 $R_\odot$, and a distance = 3.509 ± 0.009 pc, and an age of 1.33 Gyr.

**WD0839-327**

This star has a very low estimated mass and a discordant photometric and trigonometric distance. Our mass, age and luminosity is taken from GBD.

**WD0912+536**

This is magnetic WD with $B_p$ = 100 MG (see Brinkworth et al. 2013 for discussion).

**WD1008+290**

Schmidt et al. (1999) observed Zeeman splitting and a 100 MG magnetic field.

**WD1036-204**

Jordan & Friedrich (2002) find a magnetic field of 50 MG.

**WD1105-048**

The trigonometric parallax comes from the M3V companion LP672-2.

**WD1145-451**
The T$_{eff}$ and log g are from Kawka & Vennes (2012) and the BVRI photometry is from Lépine (2005).
**WD1309+853**
This is magnetic WD with B$_p$ = 4.9 MG (Putney, 1995).
**WD1310+583**
Limoges et al. (2015) gives a distance of 23.2 ± 0.8 pc, however using the same T$_{eff}$ and log g we find 26.3 ± 0.3 pc.
**WD1310-472**
Our photometric distance 14.52 ± 0.33 pc agrees with the trigonometric parallax (15.04 ± 0.54 pc) for this cool DC.
**WD1344+572**
SDSS g band is at least a magnitude too faint.
**WD1344+106**
This is magnetic WD with B$_p <$ 10 MG, Schmidt & Smith (1990).
**WD1350-090**
Our photometric distance gives an estimate of 28.40 ± 0.76 pc, while GBD finds 24.0 ± 1.1 pc and Gianninas et al. (2011) finds 20 pc. This is also a magnetic WD with B$_p <$ 10 MG, Zuckerman et al. (2003). We adopt Giammichele et al. (2012) parameters.
**WD1401+457 (J1403+4533)**
This is by far the coolest WD in our sample. Kilic et al. (2010) estimate a T$_{eff}$ of 2670 ± 1500 K and a distance of 24 pc based on an assumed log g of 8. Kilic et al. were unable to satisfactorily match the observed photometry to a model spectral energy distribution. Given the large range of T$_{eff}$ and the lack of a measured gravity, we assign a distance of 24 ± 10 pc. It is one of the lowest luminosity WDs in our sample.
**WD1532+129**
Koester et al. (2011) favour a T$_{eff}$ = 6000 ± 400 K.
**WD1620-391**
This is a well know SLS (DA1.2 + G5V) and the most luminous star in the sample (M$_{bol}$ = 7.55). This system also contains a sub-stellar companion (HD 147513b).
**WD1639+537**
This is magnetic WD with $B_p = 13$ MG, Bergeron, Ruiz & Leggett (1997, BRL) and Bergeron, Leggett, & Ruiz (2001, BLR).
**WD1658+440**
This is one of the most massive WDs known. Vennes & Kawka (2008) obtain masses of 1.33 to 1.34 M$_\odot$ and estimate a parallax 43.5 mas (23 pc). Likewise, Gianninas et al (2011) estimate a mass of 1.38 M$_\odot$ and a distance of 18 pc.
The distances are highly dependent on gravity but at a T$_{eff}$ . It is also magnetic WD with $B_p = 3.5$ MG (Schmidt et al. 1992).
**WD1820+609**
This cool DA with a weak polarization related magnetic field (B$_p \leqslant$ 0.1 MG, Putney 1997). Brinkworth et al. (2013) find evidence of long term photometric variability.
**WD1829+547**
This is magnetic WD with $B_p = 170 - 180$ MG with a rotational period > 100 yrs (Putney & Jordan 1995).
**WD1848-689**
A DC WD-M5.0 V, binary with an estimated distance of 24.6 pc, discovered by Jao et al. (2014). The cool WD (4800 K ± 200 K) is detected astrometrically in the reflex motion of the M dwarf and photometrically with HST in the 1820 Å to 2700 Å band. Astrometry yields a estimated orbital period 15.17 ± 3.5 yrs an apparent separation of 0.2082". The estimated WD mass is uncertain, but its V magnitude is estimated to be 17.4.
**WD1900+705**
This is a well studied high magnetic WD ($B_p = 100$ MG).
**WD1912+143**
This star is from Limoges, Lepine & Bergeron (2013) where it is listed as a DA7.3 high gravity DA (log g = 8.6) located at 19.4 (0.7 pc). Our photometric distance of 19.97 (1.88) is in agreement with Limoges however, the star has a trigonometric parallax form Dahn et al. (1983) of π = 28.6 ± 0.52 mas (35 pc). From the Limoges spectrum it is clearly a high gravity DA so we use the spectroscopic distance.
**WD1953-011**
This is a magnetic WD with an average global field of 70 kG. It also possesses a star spot and a rotational period of 1.4418d (see Brinkworth et al. 2005).
**WD2008-600**
This star is flux deficient in the JHK bands, due collision induced opacity (Subasavage et al. (2009).
**WD2058+342**
This is a widely separated (46") double degenerate system containing a bright DB4.1 star (GD 292A) and a fainter DC star (GD 292B). Bergeron et al. (2011) determine a distance of 24 pc for the DB star based on spectroscopic analysis. The cooler

DC star was discovered as a common proper motion companion by Farihi (2004). GD 2292A/B is an interesting system because the apparent masses of both components appear to exceed 1 M$_\odot$. Farihi's study of the DC star considers three possible distances for the system, only the nearest (25.9 pc) would seem consistent with the Bergeron et al. distance for the DB star. We tentatively leave this dd system in the 25 pc survey. It is one of the two lowest luminosity stars in our sample and its precise distance will probably be determined from parallax measurements.

**WD2105-820**
Landstreet et al. (2012) find a weak magnetic find of 8.1 to 11. 4 KG in this star.

**WD2117+539**
The *ugiz* photometry is too faint and inconsistent with the UBVRI +JHK as well as the parallax.

**WD2126+734**
This is a close (1.4") but resolved double degenerate system. Zuckerman et al. (1997) discovered the faint companion (G261-43B) and estimated its parameters. Farihi, Becklin & Zuckerman (2005) refined its parameters and provided a separation of 1.4" at a Position Angle of 167.4°. G261-43A is a DA3 star with a mass of 0.60 M$_\odot$ however its companion is 3.5 magnitudes fainter and appears to be a DC. Farihi et al. estimate a T$_{eff}$ of 6000 K and mass of 0.84 M$_\odot$. Using the orbital estimating procedures in Holberg et al. (2013) the expected physical separation of the system is 33 AU with an orbital period of 158 yrs, so that significant orbital motion ought to already be apparent since its discovery epoch. It is also noted for WD 2126+734A that the *g* and *r* band is anomalously faint compared to the V band. The companion WD 2126+734B is well defined photometrically in the SDSS data base.

**WD2133-135**
Our photometric distance is 23.81 ± 0.22 pc, while the Subasavage et al. (2008) spectroscopic fit yields a distance of 20.4 ± 3.5 pc.

**WD2140+078**
The distance is estimated from *ugriz* + JHK magnitudes as the BVRI appears inconsistent and unreliable.

**WD2140+548**
The SDSS *gri* bands are too faint.

**WD2151-015**
This is a close (< 0.5") DA6 (LTT8747A) plus M8 (LTT8747B) system. Farihi et al. (2005) provide the clearest description of this system and attempt to deconvolve the observed photometry arriving at a distance estimate of 19.4 pc. The deconvolved photometric temperature estimate of Farihi et al. is used here as well as the BVRI+JHK photometry.

**WD2226-754/755**
This is a resolved double degenerate (Subasavage et al. 2007)

**WD2307+548**
This is a DA8.8 from LBL, which appears to be a CPM with the star NLTT 56019. Newton et al. (2014) estimate a spectral type M5V at a distance of 19 pc, verses 16.2 ± 1.0 pc estimated here for the WD. The stars I band photometry is too bright.

**WD2307-691**
This is a recently recognized Sirius-Like System (Brian Skiff 2013, private communication). The primary is a K3V star with a parallax distance of 20.94 ± 0.94 pc. In the Washington Double Star Catalog (WDS J23103-6850) the secondary has an estimated V magnitude of 13.8 with separation of 13.1" at a position angle of 108°. Only V+JHK photometry is currently available but a DA star with a T$_{eff}$ =10000 K and a log g = 8.2 satisfies both the spectral energy distribution and the distance constraints.

**WD2326+049 (G29-38)**
This is the well know metal-rich, pulsating DAZH star GD 29-38. The K-band excess (Tokunaga, Becklin, & Zuckerman 1990) is clearly evident in multi-band photometric data. It also contains a weak magnetic field of B ~ 3.1 kG (see Aznar Cuadrado et al. 2005).

## Acknowledgments

We wish to acknowledge useful discussions with James Liebert and Jason Nordhous and Silvia Toonen for pointing our errors in several tables. This work was supported by NSF grant AST-1413537 to the University of Arizona and AST-1358787 to Embry-Riddle Aeronautical University. Additionally J.B.H. acknowledges support from the NASA Astrophysics Data Program grant NNX1OAD76. This research has made use of the *WD Catalog* maintained at Villanova University and the some of the data presented in this paper were obtained from the Mikulski Archive for Space Telescopes (MAST). STScI is operated by the Association of Universities for Research in Astronomy, Inc., under NASA contract NAS5-26555. Support for MAST for non-HST data is provided by the NASA Office of Space Science via grant NNX09AF08G and by other grants and contracts. Extensive use has been made of Sloan Digital Sky Survey (SDSS) photometric data. Funding for the SDSS and SDSS-II has been provided by the Alfred P. Sloan Foundation, the Participating Institutions, the National Science Foundation, the U.S. Department of Energy, the National Aeronautics and Space Administration, the Japanese

Monbukagakusho, the Max Planck Society, and the Higher Education Funding Council for England. The SDSS Web Site is http://www.sdss.org/.

The SDSS is managed by the Astrophysical Research Consortium for the Participating Institutions. The Participating Institutions are the American Museum of Natural History, Astrophysical Institute Potsdam, University of Basel, University of Cambridge, Case Western Reserve University, University of Chicago, Drexel University, Fermilab, the Institute for Advanced Study, the Japan Participation Group, Johns Hopkins University, the Joint Institute for Nuclear Astrophysics, the Kavli Institute for Particle Astrophysics and Cosmology, the Korean Scientist Group, the Chinese Academy of Sciences (LAMOST), Los Alamos National Laboratory, the Max-Planck-Institute for Astronomy (MPIA), the Max-Planck-Institute for Astrophysics (MPA), New Mexico State University, Ohio State University, University of Pittsburgh, University of Portsmouth, Princeton University, the United States Naval Observatory, and the University of Washington.

**REFERENCES**

Aznar Cuadrado R., Jordan S., Napiwotzki R., Schmid H.M., Solanki S.K., & Mathys G. 2004 A&A, 423, 1081
Bergeron P., Wesemael F., Liebert J., & Fontaine G. 1989, ApJ, 345, 91
Bergeron P., Ruiz M.-T., & Leggett S. K. 1997, ApJS, 108, 339 (BRL)
Bergeron P., Leggett S. K., & Ruiz M. T. 2001, ApJS, 133, 413 (BLR)
Bergeron P., Wesemael F., Dufour P., Beauchamp A., Hunter C., Saffer R. A., Giannias A.,
    Ruiz M. T., Limoges M.-M., Dufour P. Fontaine G., & Liebert J. 2011, ApJ, 737, 38
Bond H. E. et al. 2015, ApJ, 813, 106
Brinkworth C. S., Marsh T.R., Morales-Rueda L., Maxted P.F.L., Burleigh M.R. & Good S.A. 2005,
    MNRAS, 357, 333
Brinkworth C. S., Burleigh M. R., Lawrie K. March T. R., & Knigge C. 2013, ApJ, 773, 47
Chauvin G., Lagrange A. M., Udry S., Fusco T., Galland F., Naef D., Beuzit J. L., & Mayor M., 2006,
    A&A, 456, 1165
Crepp J. R., Johnson, J., Howard A. W., Marcy G. W., Giannias A., Kilic M., & Wright J. T. 2013,
    ApJ, 774, 1
Dahn C.C., Harrington R.S., Riepe B.Y., Christy J.W., Guetter H.H., Kallarakal V.V., Miranian M.,
    Walker R.L., Vrba F.J., Hewitt A.V., Durham W.S., & Ables H.D. 1982, AJ, 87, 419
De Rosa R. J. et al. 2013, MNRAS, 437, 1216
Dieterich S.B., Henry T.J., Golimowski D.A., Krist J.E., &Tanner A.M. 2012, AJ, 144, 64
Ducourant C., Teixeira R., Hambly N. C., Oppenheimer B. R., Hawkins M. R. S., Rapaport M.,
    Modolo J., & Lecampion J. F. 2007, A&A, 470, 387
Farihi J. 2004, ApJ, 610, 1013
Farihi J., Becklin E.E. & Zuckerman B. 2005, ApJS, 161, 394
Farihi J. 2009, MNRAS, 398, 2109
Farihi J., Burleigh M. R., Holberg J. B., Casewell S.L., & Barstow M. A. 2011, MNRAS, 417, 1735
Farihi J., Bond H. E., Dufour P. Haghighipour N., Schaefer G. H., Holberg J. B., Barstow M. A.,
    Burleigh M. R. 2013, MNRAS, 430, 652
Gatewood G., & Coban L. 2009, AJ, 137, 402
Giammichele N., Bergeron P., & Dufour P. 2012, ApJS, 199, 25 (GBD)
Giannias A., Bergeron P., & Ruiz M. T. 2011, ApJ, 743, 138
Gies D. R. et al. 2008, ApJ, 682, 117
Gould A. & Chanamé J. 2004, ApJS, 150, 455
Harrington R. S., Christy, J. W. & Strand K. Aa. 1981, AJ, 86, 909
Harris H. C. et al. 2006, AJ, 131, 571
Holberg J. B., Oswalt T. D., & Sion E. M. 2002, ApJ, 571, 512, (LWD02)
Holberg J. B., & Bergeron P. 2006, AJ, 132, 1221

Holberg J. B., Sion E. M., Oswalt, T., McCook G. P., Foran S., & Subasavage J. P. 2008, AJ, 135, 1225 (LWD08)
Holberg J. B., Bergeron P., & Gianninas A. 2008, AJ, 135, 1239 (HBG)
Holberg J.B., Oswalt T. D. Sion E. M., & Barstow M. A. 2013, MNRAS, 435, 2077
Holberg J. B. 2015, in Astronomical Society of the Pacific Conference Series, Vol. 493, 367
Jao W-C., Henry T.J., Subasavage J. P., Winters J.G., Gies, G.R., Riedell A.R., & Ianna P.A. 2014, AJ, 147, 21
Johnson K. B., Oswalt T. D., & Valls-Gabaud D. 2012, New Astronomy, 17, 458
Jordan, S., & Friedrich S. 2002, A&A, 383, 191
Kawka A., & Vennes S. 2006, ApJ, 643, 402
Kawka A., & Vennes S. 2012, MNRAS, 425, 1394
Kawka A., Vennes S., Schmidt G.D., Wickramasinghe D. T. & Koch R. 2007, ApJ, 654, 499
Kilic M., Leggett S. K., Tremblay P-E., Hippel T. V., Bergeron P., Harris H. C., Munn J. A., Williams K. A., Gates E., & Farihi J. 2010, ApJS, 190, 77
Kleinman S. J. et al. 2013, ApJS, 204, 5
Koester D., Voss B., Napiwotzki R., Christlieb N., Homeier D., Lisker T., Reimers D., & Heber U. 2009, A&A, 505, 441
Landstreet J.D., Bagulo S., Valyavin G.G., Fossati L., Jordan, S., Monin, D., & Wade G.A. 2012, A&A, 545, A30
Leggett S. K, Ruiz M.T., & Bergeron, P. 1998, ApJ, 497, 294 (LRB)
Lépine S. 2005, AJ, 130, 1237
Lépine S., Thorstensen, J. R., Shara, M. M., & Rich R. M. 2009, AJ, 137, 4109
Liebert J., et al. 2005, AJ, 129, 2376
Liebert J., Ferrario, L., Wickramasinghe D. T. & Smith, P. S. 2015, ApJ, 804, 93
Limoges M.-M., & Bergeron P. 2010, 714, 1037
Limoges M.-M., Lépine S., & Bergeron P. 2013 ApJ, 145, 136
Limoges M.-M., Bergeron P., & Lépine S. 2015, ApJS, 219, 19 (LBL)
Luyten W. J. 1979, *LHS Catalog*, 2$^{nd}$ ed. Univ. Minn. Press, p1
Mason B. D., Hartkopf W. I., Gies D. R., Henry T. J. & Helsel J. W. 2009, AJ, 137, 3358
Maxted P.F.L., O'Donoghue D., Morales-Rueda L., Napiwotzki R., & Smalley B. 2007, MNRAS, 376, 919
McCook G. P., & Sion E. M. 1999, ApJS, 121, 1 (MS99)
Nelan E. P., Bond H. E. & Schaefer G. 2015, in Astronomical Society of the Pacific Conference Series, Vol. 493, 501
Newton E. R., et al. 2014, AJ, 147, 20
Nelemans T., Yungelson L. R., Portegies Zwart S. F., & Verbunt F. 2001, A&A, 365, 49
Nordhaus J., Wellons S, Spiegel, D. S., Metzger B. D., & Blackman E. G. 2011, PNAS
Oswalt T., Strunk D., 1994, BAAS, 26, 901
Putney A. 1995, ApJ, 451, 67
Putney A., & Jordan S. 1995, ApJ, 449, 863
Putney A. 1997, ApJS, 112, 527
Raghavan D. et al. 2010, ApJS, 190, 1
Reid I. N., & Gizis J. E. 1997, AJ, 114, 1992
Reimers D., Jordan S., Koester, D., Bade N., Köhler Th. & Wisotzki,l L. 1996, A&A, 311, 572
Saffer R.A., Liebert J., & Olszewski E. W. 1988, ApJ, 334, 947
Sayres C., Subasavage J. P., Bergeron P., Dufour P., Davenport J. R. A., AlSayyar Y., & Tofflemire B. M., 2012, AJ, 143, 103
Schmidt G. D., & Smith P. S. 1995, ApJ, 448, 305
Schmidt G. D., Liebert J. & Saffer, R. A. 1992, ApJ, 394, 603

Schmidt G. D., Liebert J., Harris H. C., Dahn C. C., & Leggett S. K. 1999, ApJ, 512, 916
Sion E. M., Holberg J. B., Oswalt, T. D., McCook, G. P., & Wasatonic, R. 2009, AJ, 138, 1681
Sion E. M., Holberg J. B., Oswalt T. D., McCook G. P., Wasatonic R. & Myszka J. 2014, AJ, 147, 129
Smart R. L., Lattanzi M. G., Bucciarelli B., Massone G., Casalegno R., Chiumiento G., Drimmel, R., Lanteri L. F. Marocco F., & Spagna A. 2003, A&A, 404, 317
Subasavage J. P., Henry T. J., Bergeron P., Dufour P., Hambly N. C., & Beaulieu T. D. 2007, AJ, 134, 252
Subasavage J. P., Henry T. J., Bergeron P., Dufour P., & Hambly N. C. 2008, AJ, 136, 899
Subasavage J. P., Jao, W.-C., Henry, T. J., Bergeron P., Dufour P., Ianna P.A., Costa E. & Méndez A. 2009, AJ, 137, 4547
Tout C. A., Wickramasinghe D. T., Liebert J., Ferrario, L., & Pringle, J. E 2008, MNRAS, 387, 897
Tokunaga A. T., Becklin E. E., & Zuckerman B. 1990, ApJ, 358, L21
Tremblay P.-E., Ludwig H.-G., Steffen M., & Freytag, B. 2013, A&A, 552, 13
Tremblay P.-E., Kalirai, J. S., Soderblom D. R., Cignoni M., &Cummings J. 2014, ApJ, 791, 92
van Altena W. F., Lee J. T., & Hoffleit E. D. 1994, General Catalog of Trigonometric Stellar Parallaxes (4th ed.; New Haven, CT: Yale Univ. Observatory)
van Leeuwen F. 2007, in Astrophysics and Space Science Library Vol. 350, Hipparcos, the New Reduction of the Raw Data (Dordrecht: Springer), 20
Verbeek K. et al. 2013, MNRAS, 434, 2727
Vennes S., Thejll P.A., Galvan R.G., Dupuis J. 1997, ApJ, 480, 714
Vennes S., & Kawka A. 2008, MNRAS, 889, 1367
Zuckerman B., Becklin E. E., Macintosh B. A., & Bida T. 1997, AJ, 113, 764
Zuckerman B., Koester D., Reid I. N., & Hunsch M. 2003, ApJ, 596, 477

Table 1. 25 pc White Dwarf Sample. Column 1: The MS99 white dwarf number. Column 2: Alternate ID. Column 3: The WD spectral type . Column 4: Measured or computed V magnitude. Column 5 and 6: Observed trigonometric parallax and uncertainty. Column 7: Parallax reference. Column 8: System s = single, b = binary, m = multiple, dd = double degenerate. Column 9: Number of WDs in system. Column 10: a = stars discussed in the Appendix A.

| WD Num. | Alt | Type | V | $\pi(mas)$ | $\pi_\sigma$ (mas) | $\pi(ref)$ | Sys | Num | Appen. |
|---|---|---|---|---|---|---|---|---|---|
| WD0000-345 | LHS 1008 | DAH8.1 | 14.96 | 75.7 | 9 | 1 | s | 1 | a |
| WD0005+395 | LP 240-30 | DC10.3 | 17.12 | | | | s | 1 | |
| WD0008+424 | NLTT 529 | DA6.8 | 15.23 | | | | s | 1 | |
| WD0009+501 | LHS 1038 | DAH7.6 | 14.36 | 90.6 | 3.7 | 1 | s | 1 | a |
| WD0011-134 | LHS 1044 | DAH8.4 | 15.87 | 51.3 | 3.8 | 1 | s | 1 | a |
| WD0011-721 | NLTT 681 | DA7.8 | 15.17 | | | | s | 1 | |
| WD0025+054 | NLTT 1450 | DA9.1 | 16.31 | | | | s | 1 | |
| WD0029-031 | LHS 1093 | DA11.3 | 17.32 | 42.6 | 1 | 1 | s | 1 | |
| WD0038+555 | LTT17144 | DQ4.6 | 14.08 | 43.4 | 2 | 1 | b | 1 | |
| WD0038-226 | LHS 1126 | DQpec9.3 | 14.5 | 110.42 | 1.17 | 10 | s | 1 | |
| WD0046+051 | v Ma 2 | DZ7.4 | 12.39 | 232.7 | 1.81 | 10 | s | 1 | |
| WD0053-117 | LTT 524 | DA7.1 | 15.26 | | | | s | 1 | |
| WD0101+048 | LTT10380 | DA6.4 | 14.05 | 46.9 | 3.8 | 1 | dd | 2 | a |
| WD0108+277 | NLTT 3915 | DA9.6 | 16.15 | | | | s | 1 | a |
| WD0115+159 | LHS 1227 | DQ5.6 | 13.84 | 64.9 | 3 | 1 | s | 1 | |
| WD0121-429 | LHS 1243 | DAH7.9 | 14.83 | 54.61 | 0.96 | 3 | dd | 2 | a |
| WD0123+732 | LP 29-149 | DA6.8 | 15.61 | | | | s | 1 | a |
| WD0123-262 | L 581-26 | DC6.9 | 15.06 | | | | s | 1 | |
| WD0123-460 | SCR 0125-4545 | DA8.5 | 16.3 | | | | s | 1 | |
| WD0135-052A | LHS 1270A | DA6.9 | 12.84 | 81 | 2.8 | 1 | dd | 1 | a |
| WD0135-052B | LHS 1270B | DA6.9 | 12.84 | 81 | 2.8 | 1 | dd | 1 | a |
| WD0141-675 | LHS 145 | DA7.8 | 13.82 | 102.8 | 0.85 | 3 | s | 1 | |
| WD0145+360 | LP 244-008 | DA7.8 | 16.76 | | | | s | 1 | |
| WD0148+467 | GD 279 | DA3.8 | 12.44 | 64.51 | 3.5 | 2 | s | 1 | |
| WD0148+641 | G244-36 | DA5.6 | 13.99 | | | | b | 1 | a |
| WD0208+396 | LHS 151 | DAZ6.9 | 14.526 | 59.8 | 3.5 | 1 | s | 1 | |
| WD0210-510 | GJ86B | DA10.2 | 14 | 92.74 | 0.61 | 3 | sl | 1 | a |
| WD0213+396 | GD 25 | DA5.4 | 14.54 | | | | s | 1 | |
| WD0213+427 | LHS 153 | DA9.0 | 16.21 | 50.2 | 4.1 | 1 | s | 1 | |
| WD0227+050 | Feige 22 | DA2.7 | 12.799 | 41.15 | 4.54 | 1 | s | 1 | |
| WD0230-144 | LHS 1415 | DA9.2 | 15.77 | 64 | 3.9 | 1 | s | 1 | |
| WD0233-242 | LHS 1421 | DC9.3 | 15.94 | | | | s | 1 | |
| WD0236+259 | NLTT 8581 | DA9.2 | 16.29 | | | | s | 1 | |
| WD0243-026 | LHS 1442 | DAZ7.4 | 15.54 | 47.1 | 5 | 1 | s | 1 | |
| WD0245+541 | LHS 1446 | DAZ9.5 | 15.34 | 99.66 | 0.31 | 3 | s | 1 | |
| WD0252+497 | LP 154-64 | DA7.9 | 17.1 | | | | s | 1 | |
| WD0310-688 | LB 3303 | DA3.3 | 11.37 | 98.5 | 1.24 | 10 | s | 1 | |
| WD0311-649 | LEHPM 1-3159 | DA4.0 | 13.27 | | | | s | 1 | |
| WD0322-019 | LHS 1547 | DAZ9.9 | 16.22 | 59.5 | 3.2 | 4 | dd | 2 | a |
| WD0326-273 | LP 888-64 | DA5.4 | 13.77 | 57.6 | 13.6 | 1 | dd | 2 | a |
| WD0340+198 | LSPM | DA7.0 | 15.65 | | | | s | 1 | |

Table 1—Continued

| WD Num. | Alt | Type | V | $\pi(mas)$ | $\pi_\sigma$ (mas) | $\pi(ref)$ | Sys | Num | Appen. |
|---|---|---|---|---|---|---|---|---|---|
| WD0341+182 | LHS 179 | DQ7.7 | 15.19 | 52.6 | 3 | 1 | s | 1 | |
| WD0344+014 | LHS 5084 | DC9.9 | 16.52 | | | | s | 1 | |
| WD0357+081 | LHS 1617 | DA9.2 | 15.887 | 56.1 | 3.7 | 1 | s | 1 | |
| WD0413-077 | 40 Eri B | DA3.1 | 9.521 | 200.64 | 0.23 | 10 | sl | 1 | a |
| WD0414+420 | NLTT 12934 | DA11.0 | 17.06 | | | | s | 1 | |
| WD0416-593 | HD27442B | DA3.3 | 12.5 | 54.16 | 0.15 | 1 | sl | 1 | a |
| WD0419-487 | RR Cae | DA8 | 14.36 | 49.68 | 1.34 | 3 | b | 1 | a |
| WD0423+044 | LHS 1670 | DC9 | 17.11 | 48.26 | 3.87 | 5 | s | 1 | a |
| WD0423+120 | LB 1320 | DC8.2 | 15.42 | 57.6 | 2.5 | 4 | s | 1 | |
| WD0426+588 | G 175-34B | DC7.1 | 12.432 | 180.53 | 0.78 | 10 | b | 1 | a |
| WD0431-279 | NLTT 13532 | DC9.5 | 16.8 | | | | s | 1 | |
| WD0431-360 | LEHPM 2-1182 | DA10.0 | 17.03 | | | | s | 1 | |
| WD0433+270 | G39-027 | DA9.0 | 15.824 | 56.54 | 1.28 | 10 | sl | 1 | a |
| WD0435-088 | LHS 194 | DQ8.0 | 13.781 | 105.2 | 2.6 | 1 | s | 1 | |
| WD0454+620 | PM J | DA4.6 | 14.48 | | | | b | 1 | a |
| WD0503-174 | LHS1734 | DAH9.5 | 15.97 | 45.6 | 4 | 1 | s | 1 | |
| WD0511+079 | G84-041 | DA7.7 | 15.89 | 49.34 | 1.47 | 5 | s | 1 | |
| WD0532+414 | GD 69 | DA6.8 | 14.78 | | | | dd | 2 | |
| WD0541+260 | LSPM | DQ11.3 | 17.02 | | | | s | 1 | a |
| WD0548-001 | G99-037 | DQP8.3 | 14.58 | 90.3 | 2.8 | 1 | s | 1 | a |
| WD0552-041 | LHS 32 | DZ10.0 | 14.488 | 155.97 | 0.78 | 10 | s | 1 | |
| WD0553+053 | LHS 212 | DAP8.7 | 14.105 | 125.1 | 3.6 | 1 | s | 1 | a |
| WD0618+067 | LHS 1839 | DA8.1 | 16.37 | 44.2 | 4.2 | 1 | s | 1 | |
| WD0620-402 | LEHPM 2-505 | DZ8.5 | 16.2 | | | | s | 1 | |
| WD0628-020 | LP600-042 | DA7.2 | 15.36 | 48.84 | 1.1 | 3 | b | 1 | a |
| WD0642-166 | Sirius B | DA2 | 8.44 | 380.02 | 1.28 | 3 | sl | 1 | a |
| WD0644+025 | G108-26 | DA6.8 | 15.695 | 54.2 | 5.5 | 10 | s | 1 | |
| WD0644+375 | EG 50/LHS 1870 | DA2.4 | 12.082 | 65.51 | 1.81 | 10 | s | 1 | |
| WD0649+639 | LP 058-289 | DA8.1 | 16.28 | | | | s | 1 | |
| WD0651-398A | WT 202 | DA7.0 | 15.46 | | | | dd | 1 | a |
| WD0651-398B | WT 201 | DA8.0 | 16.07 | | | | dd | 1 | a |
| WD0655-390 | NLTT 17220 | DA7.9 | 16.22 | | | | s | 1 | a |
| WD0657+320 | LHS 1889 | DA10.1 | 16.593 | 53.5 | 0.9 | 1 | s | 1 | a |
| WD0659-063 | LHS 1892 | DA7.7 | 15.11 | 81 | 24.2 | 1 | s | 1 | a |
| WD0706+377 | G087-029 | DQ7.6 | 15.66 | 41.2 | 2.4 | 1 | b | 1 | |
| WD0708-670 | SCRJ0708-6706 | DC9.9 | 16.22 | | | 1 | s | 1 | |
| WD0727+482.1 | LHS 230A | DA10.0 | 15.26 | 90 | 1 | 1 | dd | 1 | a |
| WD0727+482.2 | LHS 230B | DA10.1 | 15.56 | 90 | 1 | 1 | dd | 1 | a |
| WD0728+642 | G234-004 | DA11.1 | 16.38 | | | | s | 1 | a |
| WD0736+053 | Procyon B | DQZ6.5 | 10.92 | 285.1 | 1.09 | 10 | sl | 1 | a |
| WD0738-172 | LHS 235 | DAZ6.6 | 13.02 | 109.94 | 0.56 | 3 | b | 1 | |
| WD0743-336 | vB03 | DC10.6 | 16.595 | 65.74 | 0.51 | 10 | sl | 1 | |
| WD0747+073.1 | LHS 239 | DC10.4 | 16.99 | 54.7 | 0.7 | 1 | dd | 1 | |
| WD0747+073.2 | LHS 240 | DC12.0 | 16.69 | 54.7 | 0.7 | 1 | dd | 1 | |
| WD0749+426 | NLTT 18555 | DC11.7 | 17.45 | | | | s | 1 | |
| WD0751-252 | SCR 0753-2524 | DA9.8 | 16.27 | 55.05 | 0.8 | 10 | b | 1 | |
| WD0752-676 | LHS 34 | DA8.8 | 14.012 | 126.62 | 1.32 | 10 | s | 1 | |
| WD0805+356 | SDSS | DA7.3 | 15.14 | | | | s | 1 | |
| WD0806-661 | L97-3 | DQ4.9 | 13.73 | 52.17 | 1.67 | 3 | s | 1 | |

Table 1—Continued

| WD Num. | Alt | Type | V | $\pi(mas)$ | $\pi_\sigma$ (mas) | $\pi(ref)$ | Sys | Num | Appen. |
|---|---|---|---|---|---|---|---|---|---|
| WD0810+489 | NLTT 19138 | DC6.9 | 15.74 | | | | s | 1 | |
| WD0816-310 | SCR 0818-3110 | DA7.8 | 15.18 | | | | s | 1 | |
| WD0821-669 | SCRJ0821-6703 | DA9.8 | 15.34 | 93.89 | 1.04 | 3 | s | 1 | |
| WD0827+328 | LHS 2022 | DA6.9 | 15.73 | 44.9 | 0.38 | 1 | s | 1 | |
| WD0839-327 | LHS 253 | DA5.5 | 11.87 | 113.59 | 1.93 | 3 | s | 1 | a |
| WD0840-136 | LP 726-1 | DZ10.3 | 15.73 | | | | s | 1 | |
| WD0856+331 | LTT 12531 | DQ5.1 | 15.18 | 48.8 | 3.4 | 1 | s | 1 | |
| WD0912+536 | PG/G049-033 | DCP7 | 13.88 | 97.3 | 1.9 | 1 | s | 1 | a |
| WD0946+534 | G195-042 | DQ6.2 | 15.3 | 43.5 | 3.5 | 1 | s | 1 | |
| WD0955+247 | LTT 12661 | DA5.8 | 15.08 | 40.9 | 4.5 | 1 | s | 1 | |
| WD0959+149 | G042-033 | DC7 | 15.37 | | | | s | 1 | |
| WD1008+290 | LS 2229 | DQpec11.0 | 17.51 | | | | s | 1 | |
| WD1009-184 | WT 1759 | DZ8.5 | 15.44 | 55.55 | 0.85 | 10 | sl | 1 | |
| WD1019+637 | LP 62-147 | DA7.2 | 14.71 | 61.2 | 3.6 | 1 | s | 1 | |
| WD1033+714 | LHS 285 | DC10.3 | 16.89 | | | | s | 1 | |
| WD1036-204 | LHS 2393 | DQpecP10.2 | 16.24 | 70 | 0.66 | 3 | s | 1 | a |
| WD1043-188 | LHS 290 | DQ8.1 | 15.51 | 56.9 | 6.5 | 1 | b | 1 | |
| WD1055-072 | LHS 2333 | DC6.8 | 14.32 | 82.3 | 3.5 | 1 | s | 1 | |
| WD1105-048 | LP 672-1 | DA3.5 | 12.92 | 57.7 | 12.5 | 1 | b | 1 | a |
| WD1116-470 | SCR | DC8.7 | 15.52 | | | | s | 1 | |
| WD1121+216 | LHS 304 | DA6.7 | 14.24 | 74.4 | 2.8 | 1 | s | 1 | |
| WD1132-325 | vB4/LHS 309 | DC10 | 15 | 104.61 | 0.37 | 10 | sl | 1 | |
| WD1134+300 | GD 140 | DA2.4 | 12.48 | 63.96 | 3.42 | 10 | s | 1 | |
| WD1142-645 | LHS 43 | DQ6.4 | 11.5 | 216.12 | 1.09 | 10 | s | 1 | |
| WD1145-451 | PM J11480-4523 | DA8.3 | 15.66 | | | | s | 1 | a |
| WD1148+687 | SDSS | DA7.6 | 15.25 | | | | s | 1 | |
| WD1149-272 | LEHPM2-4051 | DQ8.1 | 15.87 | | | | s | 1 | |
| WD1202-232 | NLTT 29555 | DAZ5.8 | 12.795 | 92.37 | 0.91 | 3 | s | 1 | |
| WD1208+076 | NLTT 29884 | DA9.6 | 16.53 | | | | s | 1 | |
| WD1208+576 | LHS 2522 | DAZ8.6 | 15.79 | 48.9 | 4.6 | 1 | s | 1 | |
| WD1223-659 | L 104-002 | DA6.6 | 13.952 | 61.53 | 1.16 | 3 | s | 1 | |
| WD1236-495 | LHS 2594 | DA4.3 | 13.767 | 61 | 9.4 | 1 | s | 1 | |
| WD1241-798 | LHS 2621 | DC/DQ | 16.18 | | | | s | 1 | |
| WD1257+037 | LHS 2661 | DA9.0 | 15.83 | 60.3 | 3.8 | 1 | sys | 1 | |
| WD1309+853 | G256-007 | DAP9 | 15.99 | 60.62 | 0.99 | 10 | s | 1 | a |
| WD1310+583 | NLTT 33303 | DA4.8 | 14.09 | | | | s | 1 | a |
| WD1310-472 | ER 8 | DC11.9 | 17.05 | 66.5 | 2.4 | 1 | s | 1 | a |
| WD1315-781 | NLTT 33551 | DC8.8 | 16.15 | 52.14 | 0.94 | 3 | s | 1 | |
| WD1327-083 | LHS 354 | DA3.6 | 12.327 | 61.55 | 2.06 | 10 | b | 1 | |
| WD1334+039 | LHS 46 | DA11 | 14.66 | 121.4 | 3.4 | 1 | s | 1 | |
| WD1336+052 | LSPM | DC12.0 | 16.65 | | | | s | 1 | |
| WD1337+705 | NLTT 34829 | DAZ2.5 | 12.773 | 40.33 | 2.68 | 1 | s | 1 | |
| WD1338+052 | LSPM | DC11.6 | 16.82 | | | | s | 1 | |
| WD1339-340 | PM J13420-3915 | DA9.5 | 16.43 | | | | s | 1 | |
| WD1344+106 | LHS 2800 | DAH7.1 | 15.08 | 49.9 | 3.6 | 1 | s | 1 | a |
| WD1344+572 | G223-024 | DA3.8 | 13.3 | | | | s | 1 | a |
| WD1345+238 | LHS 361 | DA11 | 15.65 | 82.9 | 2.2 | 1 | b | 1 | |
| WD1350-090 | LP 907-37 | DAP5 | 14.55 | | | | s | 1 | a |
| WD1401+457 | SDSS | DC19 | 18.91 | | | | s | 1 | a |

Table 1—Continued

| WD Num. | Alt | Type | V | $\pi(mas)$ | $\pi_\sigma$ (mas) | $\pi(ref)$ | Sys | Num | Appen. |
|---|---|---|---|---|---|---|---|---|---|
| WD1425-811 | BPM 784 | DAV4.2 | 13.75 | | | | s | 1 | |
| WD1436-781 | NLTT 38003 | DA8.1 | 16.11 | 40.56 | 0.8 | 3 | s | 1 | |
| WD1443+256 | LSPM | DA9.7 | 16.06 | | | | s | 1 | |
| WD1444-174 | LHS 378 | DC10.2 | 16.46 | 69 | 4 | 1 | s | 1 | |
| WD1524+297 | NLTT 40234 | DA9.9 | 17.18 | | | | s | 1 | |
| WD1532+129 | NLTT 40607 | DZ6.7 | 15.07 | | | | s | 1 | a |
| WD1544-377 | L 481-060 | DA4.8 | 12.78 | 65.14 | 0.4 | 10 | sl | 1 | |
| WD1609+135 | LHS 3163 | DA5.4 | 15.103 | 54.5 | 4.7 | 1 | s | 1 | |
| WD1620-391 | CD-38 10980 | DA2.1 | 10.977 | 78.17 | 0.38 | 10 | sl | 1 | |
| WD1625+093 | G138-031 | DA7.3 | 16.14 | 42.8 | 3.7 | 1 | s | 1 | |
| WD1626+368 | LHS 3200 | DZA6.0 | 13.834 | 62.7 | 2 | 1 | s | 1 | |
| WD1630+089 | G138-038 | DA9.0 | 15.28 | | | | s | 1 | |
| WD1632+177 | PG1632+177 | DAZ5.0 | 13.106 | | | | s | 1 | |
| WD1633+433 | LHS 6309 | DAZ7.7 | 14.834 | 66.2 | 3 | 1 | s | 1 | |
| WD1633+572 | LHS 422 | DQ8.2 | 15.004 | 69.2 | 2.5 | 1 | b | 1 | |
| WD1639+537 | LP137-043 | DAH6.7 | 15.05 | 47.4 | 3.5 | 1 | s | 1 | a |
| WD1647+591 | G226-029 | DAV4.1 | 12.21 | 90.97 | 2.31 | 10 | s | 1 | |
| WD1655+215 | LHS 3254 | DAB5.4 | 14.09 | 43 | 3.1 | 1 | s | 1 | |
| WD1658+440 | PG | DAP1.7 | 15.02 | | | | s | 1 | a |
| WD1705+030 | G139-013 | DZ7.7 | 15.194 | 57 | 5.4 | 1 | s | 1 | |
| WD1748+708 | LHS 455 | DQ9.0 | 14.15 | 164.7 | 2.4 | 1 | s | 1 | |
| WD1756+143 | LSR 17581417 | DA9.0 | 16.3 | | | | s | 1 | |
| WD1756+827 | LHS 56 | DA6.9 | 14.309 | 63.9 | 2.9 | 1 | s | 1 | |
| WD1814+134 | LSR1817+1328 | DA9.5 | 15.85 | 70.3 | 1.2 | 6 | s | 1 | |
| WD1817-598 | SCR 1821-5951 | DA5.8 | 14.77 | | | | s | 1 | |
| WD1820+609 | G227-028 | DAP10.5 | 15.67 | 78.2 | 4.1 | 1 | s | 1 | a |
| WD1829+547 | G227-035 | DXP8.0 | 15.535 | 66.8 | 5.6 | 1 | s | 1 | a |
| WD1840+042 | GD 215 | DA5.8 | 16.35 | 40.2 | 3.4 | 1 | s | 1 | |
| WD1848-689 | SCR1848 | DA10.5 | 17.4 | 40.63 | 0.72 | 7 | b | 1 | a |
| WD1900+705 | LHS 3424 | DAP4.2 | 13.23 | 77 | 2.3 | 1 | s | 1 | a |
| WD1911+536 | NLTT 47551 | DA2.9 | 13.18 | | | | s | 1 | |
| WD1912+143 | G142-020 | DA7.3 | 16.03 | 28.6 | 5.2 | 8 | s | 1 | a |
| WD1917+386 | G125-003 | DC7.9 | 14.59 | 85.5 | 3.4 | 1 | s | 1 | |
| WD1917-077 | LDS 678A | DBQZ4.9 | 12.3 | 97 | 2.12 | 10 | b | 1 | |
| WD1919+145 | GD 219 | DA3.3 | 13 | 50.5 | 5.5 | 1 | s | 1 | |
| WD1935+276 | G185-032 | DA4.2 | 12.987 | 55.7 | 2.9 | 1 | s | 1 | |
| WD1953-011 | LHS 3501 | DC6.4 | 13.698 | 87.8 | 2.9 | 1 | s | 1 | a |
| WD2002-110 | LHS 483 | DA10.5 | 16.9 | 57.7 | 0.8 | 1 | s | 1 | |
| WD2007-303 | LTT 7987 | DA3.5 | 12.18 | 61.09 | 4.51 | 2 | s | 1 | |
| WD2008-600 | SCRJ 2012-5956 | DC9.9 | 15.82 | 60.42 | 0.86 | 3 | s | 1 | a |
| WD2008-799 | SCR | DA8.5 | 16.35 | 44.7 | 1.9 | 3 | s | 1 | |
| WD2011+065 | LHS 3532 | DQ7 | 15.751 | 44.7 | 2.7 | 1 | b | 1 | |
| WD2032+248 | Wolf 1346 | DA | 11.532 | 66.98 | 1.87 | 10 | s | 1 | |
| WD2039-202 | LTT 8189 | DA2.5 | 12.34 | 47.25 | 3.44 | 10 | s | 1 | |
| WD2039-682 | BPM 13491 | DA3.1 | 13.25 | | | | s | 1 | |
| WD2040-392 | NLTT 49752 | DA4.5 | 13.74 | 44.18 | 0.97 | 3 | s | 1 | |
| WD2047+372 | G210-036 | DA3.6 | 12.97 | 57.87 | 0.67 | 3 | s | 1 | |
| WD2048+263 | LHS 3589 | DA9.7 | 15.61 | 49.8 | 3.4 | 6 | s | 1 | |

Table 1—Continued

| WD Num. | Alt | Type | V | $\pi(mas)$ | $\pi_\sigma$ (mas) | $\pi(ref)$ | Sys | Num | Appen. |
|---|---|---|---|---|---|---|---|---|---|
| WD2048-250 | NLTT 49985 | DA6.6 | 15.42 | | | | s | 1 | |
| WD2054-050 | vB11/LHS 3601 | DC10.9 | 16.68 | 59.53 | 3.11 | 10 | b | 1 | |
| WD2058+550 | LSPM | DC11.1 | 17.75 | | | | s | 1 | |
| WD2058+342A/B | GD302A | DB4.1 | 15.7 | | | | dd | 2 | |
| WD2105-820 | L 24-052 | DAP4.7 | 13.601 | 58.6 | 8.8 | 1 | s | 1 | a |
| WD2111+072 | G25-20 | DA7.8 | 16.12 | 41.1 | 3.8 | 1 | s | 1 | |
| WD2117+539 | G231-040 | DA3.6 | 12.348 | 50.7 | 7.4 | 1 | s | 1 | a |
| WD2118-388 | SCR 2122-3838 | DC9.6 | 16.55 | | | | s | 1 | |
| WD2119+040 | NLTT 51121 | DA9.0 | 16.82 | | | | s | 1 | |
| WD2126+734A | G261-43A | DA3 | 12.88 | 47.1 | 2.4 | 1 | dd | 1 | |
| WD2126+734B | G261-43B | DC10 | 12.88 | 47.1 | 2.4 | 1 | dd | 1 | a |
| WD2133-135 | NLTT 51636 | DA5.0 | 13.68 | | | 1 | s | 1 | a |
| WD2138-332 | L570-26 | DZ7 | 13.75 | 64 | 1.41 | 3 | s | 1 | |
| WD2140+078 | NLTT 51908 | DA7.7 | 15.66 | | | | s | 1 | a |
| WD2140+207 | LHS 3703 | DQ6.1 | 13.253 | 79.9 | 3.2 | 1 | s | 1 | |
| WD2149+021 | G093-048 | DA2.8 | 12.73 | 40.8 | 2.5 | 1 | s | 1 | |
| WD2151-015 | G 93-53 | DA6 | 14.515 | | | | b | 1 | a |
| WD2154-512 | LTT 8768 | DQ8.3 | 14.74 | 63 | 2.82 | 6 | b | 1 | |
| WD2159-754 | LHS 3752 | DA5.6 | 15.03 | | | | s | 1 | |
| WD2210+565 | G233-002 | DA3.0 | 12.78 | | | | m | 1 | |
| WD2211-392 | WDJ2214-390 | DA8.1 | 15.92 | 53.5 | 2.6 | 9 | s | 1 | |
| WD2215+386 | NLTT 53447 | DC10.6 | 16.99 | 39.8 | 5 | 1 | s | 1 | |
| WD2226-754 | SPMJ2231-7514 | DC11.9 | 16.59 | | | | dd | 1 | a |
| WD2226-755 | SPMJ2231-7515 | DC12.1 | 16.9 | | | | dd | 1 | a |
| WD2246+223 | LHS 3857 | DA4.7 | 14.36 | 52.5 | 4.1 | 1 | s | 1 | |
| WD2248+293 | LHS 529 | DA9 | 15.55 | 47.8 | 4.2 | 1 | s | 1 | |
| WD2251-070 | LHS 69 | DZ12.6 | 15.665 | 117.38 | 0.95 | 3 | s | 1 | |
| WD2253+054 | LTT 16738 | DA9 | 11.26 | 40.88 | 2.12 | 3 | b | 1 | |
| WD2307+548 | PM J | DA8.8 | 15.47 | | | | b | 1 | a |
| WD2307-691 | HD218572 | DA5 | 13.8 | 47.75 | 0.87 | 2 | sl | 1 | a |
| WD2311-068 | LP 702-048 | DQ6.8 | 15.42 | 39.8 | 4.7 | 1 | s | 1 | |
| WD2322+137 | NLTT 56805 | DA10.7 | 15.81 | 44.9 | 2 | 6 | s | 1 | |
| WD2326+049 | LHS 5405 | DAV4.3 | 13.05 | 73.4 | 4 | 1 | s | 1 | a |
| WD2336-079 | GD 1212 | DA4.6 | 13.26 | 62.72 | 1.7 | 3 | s | 1 | |
| WD2341+322 | G130-005 | DA4.0 | 12.932 | 56.84 | 1.78 | 10 | b | 1 | |
| WD2345+027 | NLTT 57978 | DC10.3 | 16.9 | | | | s | 1 | |
| WD2347+292 | LHS 4019 | DA9 | 15.72 | 44.65 | 4.1 | 1 | s | 1 | |
| WD2351-335 | LHS 4040 | DA5.7 | 13.56 | 43.74 | 1.43 | 3 | b | 1 | |
| WD2359-434 | LHS 1005 | DA5.9 | 12.76 | 122.41 | 1.11 | 3 | s | 1 | |

[a]Star discussed in Appendix A

References. — (1) val Altena et al. (1994), (2) van Leeuwen (2007), (3) Subasavage et al. (2009), (4) Smart et al. (2003), (5) Gatewood Coban (2009), (6) Lepine et al. (2009), (7) Joa et al. (2014), (8) Dahn et al. (1983), (9) Doucournat et al. (2007), (10) Subasavage (2013)

Table 2. UBVRI + JHK Photometry and Uncertainties

| WD Num. | U ($\sigma$) | B ($\sigma$) | V ($\sigma$) | R ($\sigma$) | I ($\sigma$) | J ($\sigma$) | H ($\sigma$) | K ($\sigma$) |
|---|---|---|---|---|---|---|---|---|
| 0000-345 | 14.98(0.09) | 15.41(0.085) | 14.96(0.085) | 14.72(0.03) | 14.45(0.03) | 14.117(0.024) | 14.024(0.038) | 13.919(0.063) |
| 0005+395 | | 17.4 | 17.12(0.03) | 16.2 | 15.8 | 15.177(0.044) | 14.85(0.066) | 14.649(0.076) |
| 0008+424 | | | 15.23(0.03) | 15 | 14.9 | 14.538(0.031) | 14.352(0.041) | 14.394(0.2) |
| 0009+501 | 14.4(0.03) | 14.82(0.03) | 14.38(0.03) | 14.08(0.03) | 13.77(0.03) | 13.49(0.022) | 13.249(0.026) | 13.191(0.03) |
| 0011-134 | 16.33(0.047) | 16.6(0.047) | 15.87(0.047) | 15.56(0.03) | 15.22(0.03) | 14.813(0.036) | 14.549(0.057) | 14.628(0.082) |
| 0011-721 | | | 15.17(0.03) | 14.87(0.03) | 14.55(0.03) | 14.211(0.034) | 13.97(0.035) | 13.915(0.049) |
| 0025+054 | | 17.2 | | 15.4 | 15.4 | 14.971(0.042) | 14.668(0.068) | 14.569(0.1) |
| 0029-031 | | | 17.32(0.03) | 16.71(0.03) | (0.03) | 15.635(0.05) | 15.38(0.091) | 15.166(0.147) |
| 0038+555 | 13.685(0.277) | 14.256(0.277) | 14.07(0.277) | | | 14.066(0.036) | 13.981(0.2) | 13.967(0.2) |
| 0038-226 | 15.05(0.047) | 15.19(0.047) | 14.52(0.047) | 14.08(0.03) | 13.71(0.03) | 13.342(0.028) | 13.483(0.033) | 13.738(0.044) |
| 0046+051 | 12.95(0.022) | 12.94(0.022) | 12.385(0.022) | 12.13(0.03) | 11.88(0.03) | 11.688(0.022) | 11.572(0.024) | 11.498(0.025) |
| 0053-117 | 15.12(0.02) | 15.605(0.02) | 15.26(0.02) | | | 14.506(0.031) | 14.348(0.2) | 14.201(0.69) |
| 0101+048 | 13.68(0.051) | 14.29(0.051) | 14.003(0.051) | 13.83(0.03) | 13.66(0.03) | 13.504(0.024) | 13.396(0.032) | 13.418(0.034) |
| 0108+277 | | 16.6 | 16.15 | 15.6 | | 15.224(0.055) | 15.042(0.084) | 14.869(0.128) |
| 0115+159 | 13.17(0.22) | 13.96(0.022) | 13.841(0.022) | 13.74(0.03) | 13.65(0.03) | 13.727(0.025) | 13.68(0.022) | 13.726(0.044) |
| 0121-429 | | | 14.83(0.03) | 14.52(0.03) | 14.19(0.03) | 13.855(0.024) | 13.634(0.038) | 13.526(0.041) |
| 0123+732 | | 16.2 | | 14.9 | 14.5 | 14.757(0.023) | 10.171(0.034) | 9.787(0.034) |
| 0123-460 | | | 16.3(0.03) | 15.94(0.03) | 15.57(0.03) | 15.111(0.041) | 14.837(0.055) | 14.906(0.102) |
| 0135-052.A | 12.702(0.03) | 13.182(0.021) | 12.83(0.02) | 12.626(0.02) | 12.371(0.02) | 12.114(0.024) | 11.954(0.022) | 11.969(0.023) |
| 0135-052.B | 12.702(0.03) | 13.182(0.021) | 12.83(0.02) | 12.626(0.02) | 12.371(0.02) | 12.114(0.024) | 11.954(0.022) | 11.969(0.023) |
| 0141-675 | 13.87(0.042) | 14.29(0.042) | 13.853(0.042) | 13.52(0.03) | 13.23(0.03) | 12.867(0.022) | 12.659(0.025) | 12.579(0.03) |
| 0145+360 | | | 16.76 | | | 15.855(0.068) | 15.478(0.113) | 15.423(0.166) |
| 0148+467 | 11.91(0.026) | 12.52(0.026) | 12.435(0.026) | | | 12.768(0.024) | 12.826(0.032) | 12.846(0.03) |
| 0148+641 | | 14.276(0.03) | 13.977(0.043) | 13.836(0.012) | 13.665(0.018) | 12.867(0.022) | 12.659(0.025) | 12.579(0.03) |
| 0208+396 | 14.36(0.03) | 14.836(0.03) | 14.514(0.02) | 14.26(0.03) | 14.04(0.03) | 13.832(0.024) | 13.67(0.034) | 13.595(0.038) |
| 0210-510 | | | 13.2(0.05) | | | | | |
| 0213+396 | 13.64(0.02) | 14.32(0.02) | 14.53(0.02) | | | 14.302(0.03) | 14.223(0.052) | 14.144(0.049) |
| 0213+427 | 16.94(0.0208) | 16.94(0.0208) | 16.213(0.0208 | 15.77(0.03) | 15.37(0.03) | 14.796(0.036) | 14.416(0.057) | 14.288(0.061) |
| 0227+050 | 11.93(0.04) | 12.76(0.04) | 12.7818(0.04) | 12.905(0.02) | 13.02(0.02) | 13.282(0.026) | 13.367(0.033) | 13.425(0.036) |
| 0230-144 | 16.475(0.03) | 16.485(0.03) | 15.765(0.03) | 15.34(0.03) | 14.93(0.03) | 14.489(0.03) | 14.261(0.048) | 14.161(0.068) |
| 0233-242 | | | 15.94(0.03) | | | 14.445(0.034) | 14.335(0.054) | 14.117(0.066) |
| 0236+259 | | 16.6 | 16.29 | | | 14.914(0.033) | 14.606(0.05) | 14.471(0.069) |
| 0243-026 | | 15.94(0.03) | 15.54(0.03) | 15.28(0.03) | 15.01(0.03) | 14.679(0.035) | 14.589(0.044) | 14.477(0.091) |
| 0245+541 | 16.69(0.106) | 16.34(0.106) | 15.414(0.106) | 14.84(0.03) | 14.36(0.03) | 13.87(0.024) | 13.545(0.04) | 13.469(0.039) |
| 0252+497 | | | 17.1 | | | 16.174(0.103) | 15.786(0.157) | 15.45(0.2) |
| 0310-688 | 10.691(0.0035 | 11.399(0.008) | 11.366(0.0016 | 11.454(0.0013 | 11.53(0.0024) | 11.758(0.023) | 11.79(0.027) | 11.861(0.023) |
| 0311-649 | | | 13.27(0.03) | 13.34(0.03) | 13.36(0.03) | 13.445(0.024) | 13.461(0.03) | 13.567(0.05) |
| 0322-019 | 17.221 | 17.02 | 16.221(0.143) | 15.66(0.03) | 15.24(0.03) | 14.761(0.042) | 14.439(0.052) | 14.378(0.084) |
| 0326-273 | | | 13.77(0.03) | 13.74 | 13.56 | 13.216(0.103) | 13.109(0.09) | 13.101(0.121) |
| 0340+198 | | | 15.65 | | | 14.949(0.031) | 14.961(0.068) | 15.038(0.124) |
| 0341+182 | 14.96(0.0106) | 15.5(0.0106) | 15.19(0.0106) | 14.91(0.03) | 14.65(0.03) | 14.59(0.031) | 14.35(0.049) | 14.23(0.06) |
| 0344+014 | | | 16.52(0.03) | 16(0.03) | 15.54(0.03) | 15(0.04) | 14.874(0.095) | 14.703(0.124) |
| 0357+081 | 16.595(0.03) | 16.595(0.03) | 15.895(0.03) | 15.49(0.03) | 15.07(0.03) | 14.562(0.038) | 14.343(0.056) | 14.122(0.057) |
| 0413-077 | 8.88(0.02) | 9.56(0.02) | 9.527(0.02) | 9.61(0.05) | 9.64(0.03) | 6.68(2) | 6.15(2) | 5.9(2) |
| 0414+420 | | 17.5 | | 16.9 | 16.4 | 15.7(0.062) | 15.363(0.088) | 15.61(0.167) |
| 0416-593 | | | 12.5 | | | 12.93(0.1) | 13.06(0.1) | 13.06(0.1) |
| 0419-487 | 14.451(0.03) | 14.901(0.03) | 14.381(0.03) | 13.76(0.03) | 12.46(0.03) | 10.72(0.024) | 10.148(0.023) | 9.852(0.025) |
| 0423+044 | | 18.12(0.03) | 17.12(0.0173) | 16.6(0.03) | 16.12(0.03) | 15.474(0.068) | 15.182(0.075) | 15.168(0.15) |
| 0423+120 | | 15.9(0.01) | 15.421(0.002) | | | 14.485(0.034) | 14.347(0.042) | 14.249(0.065) |
| 0426+588 | 12.23(0.02) | 12.75(0.0136) | 12.432(0.0136 | 12.166(0.0136 | 11.95(0.0136) | 13.721(0.116) | 12.638(0.2) | 12.451(0.2) |
| 0431-279 | | | 16.8(0.03) | 16.34(0.03) | 15.89(0.03) | 15.369(0.05) | 15.11(0.073) | 14.924(0.116) |

Table 2—Continued

| WD Num. | U (σ) | B (σ) | V (σ) | R (σ) | I (σ) | J (σ) | H (σ) | K (σ) |
|---|---|---|---|---|---|---|---|---|
| 0431-360 | | | 17.03(0.03) | 16.55(0.03) | 16.08(0.03) | 15.475(0.059) | 15.17(0.082) | 15.231(0.178) |
| 0433+270 | 16.43(0.028) | (0.028) | 15.824(0.063) | 15.4(0.03) | 15.01(0.03) | 14.598(0.038) | 14.232(0.058) | 14.136(0.069) |
| 0435-088 | 13.48(0.028) | 14.11(0.028) | 13.781(0.028) | 13.43(0.03) | 13.18(0.03) | 13.006(0.03) | 12.906(0.032) | 12.763(0.035) |
| 0454+620 | | 15.3 | 14.48(0.03) | 14.21 | 12.64 | 11.314(0.018) | 10.759(0.017) | 10.545(0.02) |
| 0503-174 | 16.72(0.03) | 16.72(0.03) | 15.99(0.03) | 15.57(0.03) | 15.11(0.03) | 14.739(0.035) | 14.408(0.047) | 14.397(0.086) |
| 0511+079 | 15.89(0.1) | 16.31(0.03) | 15.89(0.03) | | | 15.107(0.047) | 14.924(0.064) | 14.863(0.082) |
| 0532+414 | 14.59(0.07) | 15.13(0.07) | 14.81(0.07) | | | 14.087(0.029) | 13.944(0.036) | 13.816(0.049) |
| 0541+260 | | 18.1 | | 16.1 | 15.4 | 15.822(0.067) | 15.293(0.105) | 15.141(0.13) |
| 0548-001 | 14.61(0.033) | 15.06(0.033) | 14.582(0.033) | 14.22(0.03) | 13.95(0.03) | 13.73(0.029) | 13.675(0.026) | 13.705(0.043) |
| 0552-041 | 16.32(0.057) | 15.53(0.057) | 14.488(0.057) | 13.97(0.03) | 13.49(0.03) | 13.047(0.027) | 12.86(0.027) | 12.777(0.026) |
| 0553+053 | 14.57(0.0105) | 14.71(0.0105) | 14.105(0.0105) | 13.78(0.03) | 13.41(0.03) | 12.93(0.022) | 12.72(0.025) | 12.653(0.024) |
| 0618+067 | 16.78(0.03) | 16.95(0.03) | 16.41(0.03) | 16.05(0.03) | 15.67(0.03) | 15.377(0.062) | 15.017(0.071) | 14.957(0.139) |
| 0620-402 | | | 16.2(0.03) | 15.89(0.03) | 15.6(0.03) | | | |
| 0628-020 | | | 15.35(0.03) | 15.06(0.03) | 14.75(0.03) | 10.729(2) | 10.144(2) | 9.857(2) |
| 0642-166 | | | 8.44(0.05) | | | | | |
| 0644+025 | 15.35(0.0212) | 16.03(0.0212) | 15.695(0.0212) | 15.48(0.03) | 15.25(0.03) | 14.868(0.045) | 14.757(0.069) | 14.576(0.103) |
| 0644+375 | 11.09(0.027) | 12(0.027) | 12.082(0.027) | | | 12.657(0.2) | 12.662(0.061) | 12.755(0.2) |
| 0649+639 | | | 16.28 | | | 15.282(0.049) | 15.095(0.078) | 14.908(0.114) |
| 0651-398A | | | 15.46(0.03) | 15.23(0.03) | 14.98(0.03) | 14.714(0.039) | 14.549(0.052) | 14.488(0.106) |
| 0651-398B | | | 16.07(0.03) | 15.76(0.03) | 15.44(0.03) | 15.1(0.05) | 14.49(0.08) | 14.71(0.13) |
| 0655-390 | | | 15.11(0.03) | 14.81(0.03) | 14.48(0.03) | 14.15(0.033) | 13.883(0.039) | 13.888(0.069) |
| 0657+320 | 17.99(0.032) | 17.57(0.032) | 16.583(0.03) | 16.05(0.03) | 15.57(0.03) | 15.03(0.039) | 14.674(0.05) | 14.665(0.082) |
| 0659-063 | | 15.86(0.01) | 15.425(0.01) | 15.13(0.02) | 14.83(0.02) | 14.538(0.028) | 14.218(0.051) | 14.355(0.074) |
| 0706+377 | 15.37(0.04) | 15.94(0.04) | 15.66(0.039) | 15.38(0.03) | 15.16(0.03) | 15.064(0.053) | 14.783(0.051) | 14.834(0.079) |
| 0708-670 | | | 16.22(0.03) | 15.72(0.03) | 15.21(0.03) | 14.713(0.027) | 14.653(0.05) | 14.467(0.073) |
| 0727+482.1 | | 16.24(0.03) | 15.26(0.03) | 14.73(0.03) | 14.24(0.03) | 13.66(0.04) | 13.42(0.04) | 13.63(0.03) |
| 0727+482.2 | | 16.54(0.03) | 15.56(0.03) | 15.03(0.03) | 14.54(0.03) | 13.96(0.04) | 13.72(0.04) | 13.63(0.04) |
| 0728+642 | | 16.65(0.07) | 16.28(0.07) | 15.81(0.07) | | 14.811(0.033) | 14.516(0.041) | 14.385(0.066) |
| 0736+053 | | | 10.92 | | | 15.031(0.037) | 14.898(0.08) | 14.746(0.107) |
| 0738-172 | 12.67(0.033) | 13.67(0.033) | 13.024(0.033) | 12.89(0.03) | 12.72(0.03) | 12.653(0.019) | 12.611(0.026) | 12.583(0.036) |
| 0743-336 | | 17.8(0.01) | 16.595(0.01) | 15.96 | 15.39 | 14.78(0.045) | 14.548(0.055) | 14.399(0.094) |
| 0747+073.1 | | 18.17(0.03) | 16.96(0.03) | 16.31(0.03) | 15.7(0.03) | 15.031(0.037) | 14.898(0.08) | 14.746(0.107) |
| 0747+073.2 | | 17.99(0.03) | 16.93(0.03) | 16.35(0.03) | 15.22(0.03) | 14.996(0.039) | 14.719(0.067) | 14.634(0.099) |
| 0749+426 | | | 17.45 | | | 15.69(0.04) | 15.49(0.04) | 15.47(0.04) |
| 0751-252 | | | 16.27(0.03) | 15.78(0.03) | 15.31(0.03) | 14.747(0.033) | 14.471(0.033) | 14.304(0.093) |
| 0752-676 | 14.502(0.0841) | 14.665(0.084) | 14.012(0.084) | 13.58(0.03) | 13.2(0.03) | 12.726(0.023) | 12.476(0.026) | 12.362(0.024) |
| 0805+356 | | | 15.14 | | | 14.879(0.041) | 14.781(0.068) | 14.692(0.115) |
| 0806-661 | 12.86(0.04) | 13.76(0.04) | 13.73(0.03) | 13.64(0.03) | 13.6(0.03) | 13.704(0.023) | 13.739(0.025) | 13.781(0.043) |
| 0810+489 | | 15.18 | 15.74(0.03) | 14.92 | 14.97 | 14.317(0.032) | 14.132(0.042) | 14.062(0.062) |
| 0816-310 | | | 15.43(0.03) | 15.21(0.03) | 15.05(0.03) | 14.916(0.036) | 14.728(0.073) | 14.829(0.121) |
| 0821-669 | | | 15.34 | 14.82 | 14.32 | 13.792(0.029) | 13.568(0.032) | 13.339(0.039) |
| 0827+328 | 15.49(0.03) | 16.03(0.03) | 15.71(0.03) | | | 14.98(0.03) | 14.96(0.033) | 14.96(0.029) |
| 0839-327 | 11.469(0.0024) | 12.071(0.0029) | 11.846(0.0016) | 11.753(0.0013) | 11.643(0.0035) | 11.578(0.03) | 11.539(0.033) | 11.547(0.029) |
| 0840-136 | | | 15.73(0.03) | 15.36(0.03) | 15.02(0.03) | 14.623(0.032) | 14.415(0.051) | 14.543(0.087) |
| 0856+331 | 14.28(0.014) | 15.184(0.0134) | 15.174(0.0134) | 15.03(0.03) | 14.97(0.03) | 15.172(0.041) | 15.156(0.083) | 15.312(0.163) |
| 0912+536 | 13.719(0.035) | 14.226(0.035) | 13.88(0.035) | 13.629(0.02) | 13.511(0.02) | 13.308(0.025) | 13.211(0.026) | 13.133(0.03) |
| 0946+534 | 14.65(0.02) | 15.35(0.02) | 15.2(0.013) | 15.05(0.03) | 14.9(0.03) | 14.913(0.049) | 14.888(0.072) | 14.916(0.076) |
| 0955+247 | 14.97(0.03) | 15.57(0.02) | 15.08(0.03) | 14.94(0.03) | 14.54(0.03) | 14.654(0.034) | 14.659(0.069) | 14.661(0.118) |
| 0959+149 | 15.28(0.045) | 15.77(0.045) | 15.43(0.045) | | | 14.697(0.036) | 14.524(0.055) | 14.464(0.079) |
| 1008+290 | | 18.21(0.03) | 17.51(0.03) | | 15.56 | 15.125(0.046) | 14.72(0.063) | 14.535(0.077) |

Table 2—Continued

| WD Num. | U ($\sigma$) | B ($\sigma$) | V ($\sigma$) | R ($\sigma$) | I ($\sigma$) | J ($\sigma$) | H ($\sigma$) | K ($\sigma$) |
|---|---|---|---|---|---|---|---|---|
| 1009-184 | | | 15.44(0.03) | 15.18(0.03) | 14.91(0.03) | 14.681(0.036) | 14.515(0.055) | 14.314(0.072) |
| 1019+637 | 14.572(0.01) | 15.082(0.01) | 14.71(0.007) | 14.45(0.03) | 14.17(0.03) | 13.874(0.029) | 13.733(0.047) | 13.692(0.049) |
| 1033+714 | | 17.92(0.021) | 16.89(0.021) | 16.33(0.03) | 15.8(0.03) | 15.194(0.048) | 14.839(0.084) | 14.98(0.148) |
| 1036-204 | | 16.37 | 16.24(0.03) | 15.54(0.03) | 15.34(0.03) | 14.633(0.033) | 14.346(0.041) | 14.035(0.067) |
| 1043-188 | | 16.015(0.021) | 15.51(0.021) | 15.03(0.03) | | 14.62(0.019) | 14.41(0.026) | 14.36(0.026) |
| 1055-072 | 14.098(0.036) | 14.636(0.036) | 14.32(0.036) | 14.13(0.03) | 13.91(0.03) | 13.77(0.029) | 13.68(0.032) | 13.485(0.038) |
| 1105-048 | 12.42(0.02) | 13.11(0.02) | 13.057(0.02) | 13.189(0.003) | 13.26(0.002) | 13.405(0.026) | 13.445(0.03) | 13.544(0.056) |
| 1116-470 | | | 15.52(0.03) | 15.2(0.03) | 14.86(0.03) | 14.454(0.032) | 14.368(0.055) | 14.351(0.092) |
| 1121+216 | 14(0.03) | 14.53(0.03) | 14.237(0.03) | 14.01(0.03) | 13.76(0.03) | 13.574(0.024) | 13.42(0.026) | 13.399(0.034) |
| 1132-325 | | | 15 | | | 13.56(0.1) | 13.44(0.1) | 13.43(0.1) |
| 1134+300 | 11.468(0.029) | 12.407(0.008) | 12.493(0.016) | 12.605(0.007) | 12.722(0.019) | 12.993(0.024) | 13.105(0.031) | 13.183(0.028) |
| 1142-645 | 11.063(0.033) | 11.683(0.033) | 11.499(0.033) | 11.33(0.02) | 11.17(0.02) | 11.188(0.024) | 11.13(0.025) | 11.104(0.026) |
| 1145-451 | | 16.9(0.3) | 15.66(0.25) | 14.2(0.03) | 13.8(0.3) | 14.888(0.38) | 14.762(0.068) | 14.596(0.83) |
| 1148+687 | | | | | | 14.462(0.034) | 14.22(0.04) | 14.218(0.056) |
| 1149-272 | | | 15.87(0.03) | 15.59(0.03) | 15.37(0.03) | 15.169(0.046) | 14.921(0.065) | 14.769(0.11) |
| 1151+246 | | 15.6 | 15.26 | | | 15.236(0.037) | 15.109(0.067) | 15.405(0.151) |
| 1202-232 | 12.485(0.01) | 13.045(0.01) | 12.795(0.007) | 12.66(0.03) | 12.52(0.03) | 12.402(0.024) | 12.301(0.027) | 12.342(0.026) |
| 1208+076 | | 17.2 | | 16.1 | 16.4 | 15.373(0.063) | 15.087(0.071) | 15.1(0.044) |
| 1208+576 | 16.14(0.01) | 16.35(0.01) | 15.787(0.01) | 15.41 | 15.04 | 14.679(0.034) | 14.362(0.052) | 14.458(0.095) |
| 1223-659 | 13.812(0.016) | 14.352(0.016) | 13.952(0.016) | 13.82(0.03) | 13.62(0.03) | 13.334(0.041) | 13.257(0.051) | 13.297(0.056) |
| 1236-495 | 13.26(0.006) | 13.957(0.0044 | 13.782(0.0041 | 13.775(0.0035 | 13.748(0.0136 | 13.806(0.024) | 13.815(0.036) | 13.907(0.062) |
| 1241-798 | | | 16.18(0.03) | 15.8(0.03) | 15.45(0.03) | 15.034(0.048) | 14.834(0.068) | 14.603(0.119) |
| 1257+037 | 16.38(0.053) | 16.447(0.053) | 15.832(0.053) | 15.38(0.02) | 15.01(0.02) | 14.655(0.04) | 14.316(0.05) | 14.22(0.089) |
| 1309+853 | 16.98(0.021) | 16.765(0.021) | 15.985(0.021) | | | 14.686(0.034) | 14.459(0.063) | 14.342(0.09) |
| 1310+583 | | | 14.09 | | | 14.016(0.0028 | 14(0.0045) | 14.081(0.073) |
| 1310-472 | | 18.353(0.046) | 17.077(0.046) | 16.41(0.03) | 15.72(0.03) | 15.13(0.04) | 15.04(0.08) | 14.73(0.12) |
| 1315-781 | | | 16.16(0.03) | 15.73(0.03) | 15.35(0.03) | 14.888(0.044) | 14.666(0.077) | 14.58(0.118) |
| 1327-083 | 11.712(0.019) | 12.322(0.019) | 12.327(0.019) | 12.45(0.02) | 12.45(0.03) | 12.621(0.037) | 12.677(0.041) | 12.736(0.048) |
| 1334+039 | 15.952(0.033) | 15.615(0.033) | 14.66(0.033) | 14.084(0.02) | 13.63(0.03) | 13.064(0.024) | 12.819(0.026) | 12.69(0.021) |
| 1336+052 | | 17.5 | | 16 | 15.5 | 14.699(0.024) | 14.592(0.051) | 14.478(0.073) |
| 1337+705 | 11.807(0.0022 | 12.682(0.0017 | 12.77(0.003) | 12.873(0.0013) | 12.979(0.002) | 13.248(0.024) | 13.357(0.027) | 13.451(0.035) |
| 1338+052 | | | 16.82 | | | 14.699(0.024) | 14.592(0.051) | 14.478(0.073) |
| 1339-340 | | | 16.43(0.03) | 16(0.03) | 15.56(0.03) | 15(0.044) | 14.747(0.063) | 14.651(0.095) |
| 1344+106 | 14.99(0.03) | 15.48(0.03) | 15.12(0.03) | 14.9(0.03) | 14.64(0.03) | 14.407(0.038) | 14.139(0.053) | 14.235(0.08) |
| 1344+572 | | | 13.3 | | | 13.704(0.022) | 13.821(0.035) | 13.757(0.041) |
| 1345+238 | 17.232(0.032) | 16.762(0.032) | 15.654(0.032) | 15.12(0.03) | 14.58(0.03) | 13.921(0.027) | 13.669(0.036) | 13.621(0.04) |
| 1350-090 | | | 14.55 | | | 14.208(0.2) | 14.145(0.054) | 14.139(0.2) |
| 1401+457 | | 19.37 | 18.81 | | | | | |
| 1425-811 | 13.185(0.05) | 13.51(0.05) | 13.445(0.058) | | | 13.577(0.022) | 13.603(0.025) | 13.645(0.047) |
| 1436-781 | | | 16.1(0.03) | 15.81(0.03) | 15.48(0.03) | 15.039(0.04) | 14.88(0.082) | 14.758(0.139) |
| 1443+256 | | 16.8 | | 15.5 | 15.3 | 14.676(0.035) | 14.132(0.045) | 14.206(0.082) |
| 1444-174 | | 17.445(0.03) | 16.46(0.03) | 15.97(0.028) | 15.45(0.03) | 14.948(0.029) | 14.64(0.047) | 14.724(0.108) |
| 1524+297 | | 18.4 | | 16.1 | 15.8 | 15.3(0.056) | 14.965(0.084) | 14.8(0.144) |
| 1532+129 | | | 15.07(0.03) | | | 14.939(0.036) | 14.728(0.045) | 14.706(0.08) |
| 1538+333 | | | 15.03 | | | 14.772(0.034) | 14.703(0.057) | 14.494(0.063) |
| 1544-377 | 12.69(0.047) | 13.116(0.047) | 12.783(0.047) | 12.75(0.007) | 12.37(0.007) | 12.94(0.1) | 12.84(0.1) | 12.85(0.1) |
| 1609+135 | 14.676(0.03) | 15.321(0.02) | 15.098(0.01) | 15.01(0.03) | 14.87(0.03) | 14.861(0.036) | 14.779(0.056) | 14.857(0.109) |
| 1620-391 | 9.883(0.0019) | 10.881(0.0019) | 11.008(0.002) | 11.135(0.0013) | 11.274(0.0025) | 11.577(0.021) | 11.708(0.023) | 11.768(0.022) |
| 1625+093 | 16.008(0.006) | 16.475(0.007) | 16.117(0.007) | 15.899(0.007) | 15.68(0.0128) | 15.25(0.062) | 15.187(0.103) | 15.036(0.142) |
| 1626+368 | 13.343(0.0232 | 14.008(0.0232 | 13.834(0.0232 | 13.75(0.03) | 13.66(0.03) | 13.637(0.024) | 13.652(0.034) | 13.575(0.042) |

Table 2—Continued

| WD Num. | U (σ) | B (σ) | V (σ) | R (σ) | I (σ) | J (σ) | H (σ) | K (σ) |
|---|---|---|---|---|---|---|---|---|
| 1630+089 | | | 15.28 | | | 13.849(0.026) | 13.611(0.029) | 13.488(0.033) |
| 1632+177 | | | 13.106 | | | 13.049(0.021) | 12.989(0.027) | 13.076(0.029) |
| 1633+433 | 14.835(0.02) | 15.255(0.02) | 14.83(0.02) | 14.57(0.03) | 14.28(0.03) | 13.991(0.029) | 13.773(0.035) | 13.607(0.043) |
| 1633+572 | 15.134(0.0166 | 15.494(0.0166 | 15.004(0.0166 | 15.11(0.03) | 14.28(0.03) | 14.11(0.028) | 14.077(0.037) | 14.135(0.062) |
| 1639+537 | 14.763(0.013) | 15.349(0.013) | 15.044(0.013) | 14.766(0.018) | 14.57(0.03) | 14.493(0.027) | 14.479(0.048) | 14.369(0.085) |
| 1647+591 | 11.754(0.0339 | 12.389(0.0339 | 12.214(0.0339 | | | 12.425(0.021) | 12.463(0.021) | 12.522(0.03) |
| 1655+215 | 13.739(0.04) | 14.335(0.03) | 14.102(0.025) | 14.03(0.03) | 13.92(0.03) | 13.886(0.026) | 13.816(0.03) | 13.863(0.05) |
| 1658+440 | | | 14.62(0.03) | | | 15.445(0.056) | 15.526(0.128) | 15.048(0.165) |
| 1705+030 | 15.403(0.0154 | 15.649(0.0154 | 15.194(0.0154 | 14.96(0.03) | 14.74(0.03) | 14.565(0.032) | 14.499(0.032) | 14.511(0.078) |
| 1748+708 | 14.277(0.0173 | 14.577(0.0173 | 14.151(0.0173 | 13.6(0.02) | 13.08(0.02) | 12.709(0.021) | 12.528(0.023) | 12.507(0.023) |
| 1756+143 | | | 16.3(0.03) | 16.12(0.03) | 15.69(0.03) | 14.931(0.041) | 14.659(0.061) | 14.659(0.079) |
| 1756+827 | 14.144(0.0431 | 14.663(0.0431 | 14.309(0.0431 | 14.12(0.03) | 13.87(0.03) | 13.634(0.025) | 13.472(0.03) | 13.431(0.05) |
| 1814+134 | | | 15.85(0.03) | 15.34(0.03) | 14.86(0.03) | 14.377(0.036) | 14.1(0.057) | 14.075(0.057) |
| 1817-598 | | | 16.85(0.03) | 16.3(0.03) | 15.8(0.03) | 15.22(0.05) | 14.98(0.05) | 14.89(0.05) |
| 1820+609 | | 16.645(0.03) | 15.67(0.0283) | 15.33(0.03) | 14.62(0.03) | 14.075(0.032) | 13.81(0.03) | 13.797(0.052) |
| 1829+547 | 15.585(0.0495 | 16.025(0.0495 | 15.535(0.0495 | 15.28(0.03) | 14.97(0.03) | 14.803(0.045) | 14.478(0.053) | 14.505(0.078) |
| 1840+042 | 14.427(0.061) | 15.032(0.061) | 14.797(0.061) | | | 14.443(0.05) | 14.374(0.075) | 14.651(0.099) |
| 1848-689 | | | 17.4 | | | | | |
| 1900+705 | 12.423(0.323) | 13.276(0.03) | 13.226(0.03) | 13.24(0.03) | 13.23(0.03) | 13.334(0.023) | 13.439(0.031) | 13.417(0.048) |
| 1911+536 | | | 13.18 | | | 13.616(0.029) | 13.734(0.046) | 13.824(0.047) |
| 1912+143 | | | 16.03 | | | 15.26(0.08) | 14.92(0.1) | 14.78(0.1) |
| 1917+386 | 14.657(0.054) | 15.031(0.054) | 14.586(0.054) | 14.31 | 14.04 | 13.776(0.03) | 13.669(0.032) | 13.519(0.025) |
| 1917-077 | 11.515(0.0488 | 12.349(0.0488 | 12.297(0.0488 | 10.65(0.02) | 9.41(0.02) | 12.351(0.026) | 12.355(0.025) | 12.421(0.026) |
| 1919+145 | 12.411(0.0505 | 13.071(0.0505 | 13.001(0.0505 | | | 13.261(0.2) | 13.452(0.055) | 13.546(0.066) |
| 1935+276 | 12.594(0.0201 | 13.157(0.0201 | 12.987(0.0201 | | | 13.183(0.023) | 13.213(0.029) | 13.329(0.043) |
| 1953-011 | 13.381(0.0264 | 13.971(0.0264 | 13.698(0.0264 | 13.5(0.02) | 13.31(0.02) | 13.07(0.029) | 13.029(0.031) | 13.014(0.04) |
| 2002-110 | | 17.977(0.0462 | 16.897(0.0462 | 16.36(0.03) | 15.86(0.03) | 15.276(0.055) | 14.995(0.072) | 14.746(0.105) |
| 2007-303 | 11.576(0.006) | 12.269(0.0042 | 12.206(0.0021 | 12.352(0.0024 | 12.418(0.0069 | 12.583(0.024) | 12.641(0.027) | 12.698(0.032) |
| 2008-600 | | | 15.84(0.03) | 15.4(0.03) | 14.99(0.03) | 14.928(0.045) | 15.25(0.05) | 15.41(0.05) |
| 2008-799 | | | 16.35(0.03) | 15.96(0.03) | 15.57(0.03) | 15.108(0.04) | 15.027(0.081) | 14.643(0.09) |
| 2011+065 | 15.733(0.012) | 16.176(0.067) | 15.751(0.0129 | 15.476(0.009) | 15.25(0.016) | 15.021(0.049) | 14.878(0.2) | 15.09(0.2) |
| 2032+248 | 10.592(0.013) | 11.46(0.013) | 11.523(0.013) | | | 12.039(0.026) | 12.072(0.032) | 12.186(0.027) |
| 2039-202 | 11.446(0.03) | 12.286(0.03) | 12.4(0.02) | 12.5(0.02) | 12.62(0.02) | 12.825(0.028) | 12.921(0.027) | 13.004(0.03) |
| 2039-682 | 12.769(0.11) | 13.549(0.11) | 13.439(0.11) | | | 13.729(0.026) | 13.806(0.039) | 13.8(0.05) |
| 2040-392 | | | 13.75(0.03) | 13.76(0.03) | 13.69(0.03) | 13.775(0.024) | 13.822(0.034) | 13.81(0.049) |
| 2047+372 | 12.436(0.05) | 13.106(0.05) | 12.966(0.05) | | | 13.303(0.023) | 13.37(0.024) | 13.432(0.038) |
| 2048+263 | 16.563 | 16.163(0.025) | 15.613(0.025) | 15.11(0.03) | 14.63(0.03) | 14.1(0.056) | 13.908(0.068) | 13.602(0.2) |
| 2048-250 | | 15.8(0.024) | 15.42(0.024) | 15.2(0.03) | | 14.9(0.037) | 14.704(0.045) | 14.599(0.083) |
| 2054-050 | 18.45(0.03) | 17.81(0.03) | 16.64(0.03) | 15.98(0.03) | 15.32(0.03) | 14.734(0.081) | 14.565(0.134) | 14.327(0.136) |
| 2058+550 | | 18.7 | | 17 | 16.4 | 15.663(0.063) | 15.449(0.108) | 15.492(0.199) |
| 2058+342 | | 15.744(0.04) | 15.674(0.04) | | | 15.744(0.07) | 15.848(0.155) | 15.568(0.193) |
| 2105-820 | 13.224(0.02) | 13.83(0.03) | 13.596(0.03) | 13.56(0.03) | 13.47(0.03) | 13.478(0.026) | 13.451(0.033) | 13.533(0.039) |
| 2110+072 | | | 16.12 | | | 15.21(0.05) | 14.95(0.08) | 14.81(0.1) |
| 2115-560 | 13.955(0.01) | 14.545(0.01) | 14.285(0.01) | | | 14.11(0.029) | 13.996(0.055) | 14.022(0.061) |
| 2117+539 | 11.748(0.023) | 12.415(0.023) | 12.348(0.023) | | | 12.681(0.021) | 12.785(0.023) | 12.85(0.038) |
| 2118-388 | | | 16.55(0.03) | 16.09(0.03) | 15.65(0.03) | 15.163(0.043) | 14.916(0.067) | 15.049(0.119) |
| 2119+040 | | | 16.82 | | | 15.244(0.053) | 15.009(0.07) | 14.876(0.106) |
| 2126+734A | 12.173(0.043) | 12.859(0.043) | 12.829(0.043) | | | 13.096(0.03) | 13.164(0.038) | 13.166(0.044) |
| 2126+734B | | | | | 15.7 | 15.34 | 15.05 | |
| 2133-135 | | | 13.68(0.03) | 13.63(0.03) | 13.55(0.03) | 13.604(0.032) | 13.576(0.036) | 13.687(0.058) |

Table 2—Continued

| WD Num. | U ($\sigma$) | B ($\sigma$) | V ($\sigma$) | R ($\sigma$) | I ($\sigma$) | J ($\sigma$) | H ($\sigma$) | K ($\sigma$) |
|---|---|---|---|---|---|---|---|---|
| 2138-332 | | | 14.47(0.03) | 14.3(0.03) | 14.16(0.03) | 14.167(0.031) | 14.078(0.043) | 13.946(0.061) |
| 2140+078 | | 18 | 15.6(0.022) | 15.5 | 13.65 | 15.774(0.081) | 15.271(0.109) | 15.178(0.156) |
| 2140+207 | 12.708(0.022) | 13.418(0.022) | 13.253(0.022) | 13.1(0.03) | 12.98(0.03) | 12.981(0.021) | 12.928(0.035) | 12.922(0.029) |
| 2149+021 | 12.019(0.048) | 12.785(0.048) | 12.75(0.02) | 12.85(0.02) | 12.55(0.02) | 13.203(0.024) | 13.286(0.037) | 13.397(0.037) |
| 2151-015 | | 14.728(0.018) | 14.515(0.022) | 14.373(0.013) | 14.24(0.1) | 14.16(0.1) | 14.06(0.1) | 14.09(0.1) |
| 2154-512 | 14.116(0.061) | 14.916(0.061) | 14.741(0.061) | 14.3(0.03) | 14.13(0.03) | 14.007(0.032) | 13.908(0.028) | 13.806(0.046) |
| 2159-754 | 14.561(0.041) | 15.191(0.041) | 15.03(0.041) | | | 14.722(0.039) | 14.674(0.067) | 14.547(0.099) |
| 2210+565 | | | 12.75 | | | 13.191(0.039) | 12.986(0.04) | 12.859(0.036) |
| 2211-392 | | 16.41 | 15.92(0.03) | 15.61(0.03) | 15.24(0.03) | 14.892(0.033) | 14.644(0.054) | 14.557(0.077) |
| 2215+386 | | | 16.99(0.03) | | | 15.405(0.053) | 15.196(0.095) | 14.97(0.142) |
| 2226-754 | | | 16.58(0.03) | 15.92(0.03) | 15.32(0.03) | 14.662(0.037) | 14.658(0.058) | 14.436(0.079) |
| 2226-755 | | | 16.89(0.03) | 16.16(0.03) | 15.51(0.03) | 14.858(0.039) | 14.824(0.06) | 14.723(0.12) |
| 2246+223 | 13.861(0.02) | 14.548(0.02) | 14.358(0.02) | 14.34(0.03) | 14.25(0.03) | 14.341(0.029) | 14.317(0.047) | 14.36(0.09) |
| 2248+293 | 16.06(0.03) | 16.178(0.03) | 15.54(0.03) | 15.14(0.03) | 14.75(0.03) | 14.316(0.029) | 13.983(0.038) | 13.941(0.044) |
| 2251-070 | | 17.555(0.03) | 15.665(0.03) | 15.1(0.03) | 14.56(0.03) | 14.341(0.029) | 14.317(0.047) | 14.36(0.09) |
| 2253+054 | | 16.717(0.014) | 16.174(0.013) | 15.789(0.012) | 15.27(0.017) | 15.182(0.099) | 14.886(0.126) | 15.02(0.178) |
| 2311-068 | 14.978(0.039) | 15.614(0.039) | 15.404(0.04) | 15.25(0.02) | 15.1(0.02) | 14.951(0.036) | 14.942(0.071) | 14.73(0.093) |
| 2322+137 | | | 15.81(0.03) | | | 14.512(0.034) | 14.371(0.049) | 14.364(0.095) |
| 2307+548 | | 16 | | 14.3 | 11.4 | 14.269(0.051) | 13.941(0.075) | 13.93(0.073) |
| 2307-691 | | | 13.81 | | | 13.595(0.058) | 13.639(0.12) | 13.664(0.096) |
| 2326+049 | 12.583(0.02) | 13.216(0.02) | 13.03(0.03) | 13.03(0.02) | 13.01(0.02) | 13.132(0.029) | 13.075(0.022) | 12.689(2) |
| 2336-079 | | | 13.28(0.03) | 13.27(0.03) | 13.24(0.03) | 13.339(0.029) | 13.341(0.023) | 13.35(0.03) |
| 2341+322 | 12.477(0.051) | 13.068(0.051) | 12.932(0.051) | | | 13.171(0.029) | 13.195(0.037) | 13.179(0.028) |
| 2345+027 | | 17.8 | | 16.2 | 15.5 | 15.615(0.064) | 15.596(0.137) | 15..867(0.072) |
| 2347+292 | 16.08(0.03) | 16.322(0.03) | 15.737(0.02) | 15.41(0.03) | 15.04(0.03) | 14.571(0.029) | 14.345(0.044) | 14.159(0.065) |
| 2351-335 | | | 14.52(0.011) | 14.38(0.03) | 14.19(0.106) | 13.985(0.106) | 13.855(0.249) | 13.725(0.113) |
| 2359-434 | 11.971(0.196) | 12.841(0.196) | 12.911(0.196) | 12.815(0.02) | 12.647(0.02) | 12.597(0.026) | 12.425(0.023) | 12.445(0.024) |

Table 3. Physical Properties Column 1: The MS99 white dwarf number. Column 2: Adopted effective temperature and unc. Column 3: Adopted log surface gravity and unc. Column 4: Photospheric Composition. Column 5: Estimated Mass and uncertainty in solar masses. Column 6: Distance and unc in pc. Column 7: Method by which distance was determined(p = parallax and s = spectroscopic). Column 8: Estimated bolometric magnitude. Column 9: Estimated cooling age in Gyr.

| WD Num. | $T_{eff}(K)(\sigma)$ | log g ($\sigma$) | Comp | $M_\odot$ ($\sigma$) | Dist. (pc)($\sigma$) | Method | $M_{bol}$ | Age(Gyr) |
|---|---|---|---|---|---|---|---|---|
| 0000-345 | 6644( 149) | 8.42(0.16) | H | 0.86(0.10) | 13.21(1.57) | p | 14.27 | 3.84 |
| 0005+395 | 4890(435) | 8.00(0.00) | H | 0.58(0.00) | 20.21(1.25) | s | 14.97 | 6.14 |
| 0008+424 | 7203( 107) | 7.99(0.07) | H | 0.59(0.03) | 23.64(0.50) | s | 13.25 | 1.4 |
| 0009+501 | 6503( 143) | 8.21(0.05) | H | 0.72(0.03) | 11.04(0.45) | p | 14.03 | 2.89 |
| 0011-134 | 5992( 114) | 8.21(0.11) | H | 0.72(0.07) | 19.49(1.44) | p | 14.39 | 3.52 |
| 0011-721 | 6326( 152) | 7.98(0.00) | H | 0.58(0.00) | 17.84(0.24) | s | 13.8 | 1.93 |
| 0025+054 | 5520(855) | 8.00(0.00) | H | 0.58(0.01) | 21.12(1.71) | s | 14.43 | 3.13 |
| 0029-031 | 4470( 50) | 8.07(0.04) | H | 0.62(0.03) | 23.47(0.55) | p | 15.47 | 7.96 |
| 0038+555 | 10632( 630) | 7.95(0.07) | H | 0.58(0.03) | 23.04(1.06) | p | 11.48 | 0.49 |
| 0038-226 | 5229( 76) | 7.93(0.02) | H+He | 0.53(0.01) | 9.06(0.10) | p | 14.59 | 4.44 |
| 0046+051 | 6216( 183) | 8.15(0.03) | He | 0.68(0.02) | 4.30(0.03) | p | 14.17 | 3.3 |
| 0053-117 | 7146( 108) | 8.05(0.08) | H | 0.62(0.05) | 22.22(0.68) | s | 13.37 | 1.65 |
| 0101+048 | 8039( 200) | 7.33(0.14) | H | 0.30(0.04) | 21.32(1.73) | p | 11.89 | 0.53 |
| 0108+277 | 5270( 250) | 8.36(0.60) | H | 0.82(0.39) | 14.88(2.00) | s | 15.19 | 6.93 |
| 0115+159 | 9061( 319) | 7.90(0.07) | He | 0.69(0.04) | 15.41(0.71) | p | 12.49 | 1.04 |
| 0121-429 | 6230( 191) | 7.64(0.03) | H | 0.41(0.02) | 18.31(0.32) | p | 13.42 | 1.38 |
| 0123+732 | 7400(11) | 8.20(0.10) | H | 0.66(0.06) | 25.10(1.40) | s | 13.3 | 1.72 |
| 0123-262 | 7290( 192) | 7.91(0.00) | He | 0.58(0.00) | 21.7(0.8) | s | 13.24 | 1.47 |
| 0123-460 | 5898( 161) | 8.00(0.00) | H | 0.59(0.00) | 25.30(0.46) | s | 14.14 | 2.42 |
| 0135-052A | 7470(500) | 7.80(0.10) | H | 0.43(0.03) | 12.35(0.43) | p | 12.67 | 1.12 |
| 0135-052B | 6290(500) | 7.89(0.10) | H | 0.52(0.05) | 12.35(0.43) | p | 13.68 | 1.76 |
| 0141-675 | 6249( 99) | 7.79(0.10) | H | 0.48(0.05) | 9.73(0.08) | p | 13.61 | 1.65 |
| 0145+360 | 6471( 138) | 8.63(0.28) | H | 0.99(0.16) | 23.25(2.92) | s | 14.75 | 4.47 |
| 0148+467 | 14005( 277) | 8.04(0.04) | H | 0.64(0.03) | 15.50(0.84) | p | 10.41 | 0.27 |
| 0148+641 | 8963( 129) | 7.84(0.05) | H | 0.52(0.02) | 16.99(0.20) | s | 12.08 | 0.69 |
| 0208+396 | 7266( 176) | 7.91(0.09) | H | 0.55(0.04) | 16.72(0.98) | p | 13.1 | 1.28 |
| 0210-510 | 8184( 120) | 7.81(0.02) | H | 0.50(0.01) | 10.78(0.04) | p | 12.45 | 0.86 |
| 0213+396 | 9305( 135) | 8.05(0.05) | H | 0.63(0.03) | 22.99(0.41) | s | 12.21 | 0.83 |
| 0213+427 | 5507( 108) | 8.10(0.13) | H | 0.65(0.09) | 19.92(1.63) | p | 14.59 | 3.93 |
| 0227+050 | 19920( 305) | 7.93(0.05) | H | 0.59(0.02) | 24.30(2.68) | p | 8.68 | 0.07 |
| 0230-144 | 5528( 15) | 8.15(0.19) | H | 0.68(0.12) | 15.63(0.95) | p | 14.66 | 4.5 |
| 0233-242 | 5312( 146) | 8.00(0.00) | H | 0.58(0.00) | 16.83(0.36) | s | 14.6 | 4.09 |
| 0236+259 | 5666( 114) | 8.00(0.00) | H | 0.59(0.00) | 21.01(0.00) | s | 14.32 | 2.85 |
| 0243-026 | 6771( 156) | 8.14(0.16) | H | 0.68(0.10) | 21.23(2.25) | p | 13.75 | 2.33 |
| 0245+541 | 5139( 82) | 8.23(0.05) | H | 0.73(0.03) | 10.03(0.03) | p | 15.1 | 6.51 |
| 0252+497 | 6370( 166) | 9.02(0.31) | H | 1.20(0.11) | 17.99(2.90) | s | 15.58 | 4.51 |
| 0310-688 | 16865( 244) | 8.10(0.04) | H | 0.68(0.02) | 10.15(0.13) | p | 9.68 | 0.17 |
| 0311-649 | 11654( 557) | 7.95(0.00) | H | 0.58(0.00) | 21.58(0.28) | s | 11.08 | 0.39 |
| 0322-019 | 5195( 87) | 8.08(0.08) | H | 0.63(0.05) | 16.81(0.90) | p | 14.82 | 5.24 |
| 0326-273 | 8438( 200) | 7.48(0.39) | H | 0.34(0.11) | 17.36(4.10) | p | 11.86 | 0.54 |
| 0340+198 | 7161( 47) | 8.55(0.09) | H | 0.94(0.05) | 19.50(0.83) | s | 14.16 | 3.65 |

Table 3—Continued

| WD Num. | $T_{eff}(K)(\sigma)$ | log g ($\sigma$) | Comp | $M_\odot$ ($\sigma$) | Dist. (pc)($\sigma$) | Method | $M_{bol}$ | Age(Gyr) |
|---|---|---|---|---|---|---|---|---|
| 0341+182 | 6569( 136) | 7.96(0.09) | He | 0.57(0.06) | 19.01(1.08) | p | 13.67 | 1.68 |
| 0344+014 | 5170( 171) | 8.00(0.00) | H | 0.58(0.00) | 20.78(0.52) | s | 14.72 | 4.81 |
| 0357+081 | 5565( 18) | 8.05(0.12) | H | 0.62(0.08) | 17.83(1.18) | p | 14.47 | 3.42 |
| 0413-077 | 17100( 256) | 7.95(0.04) | H | 0.51(0.036 | 4.98(0.01) | p | 9.39 | 0.12 |
| 0414+420 | 4580(374) | 8.00(0.00) | H | 0.58(0.00) | 23.80(3.60) | s | 15.25 | 7.31 |
| 0416-593 | 15310( 358) | 7.98(0.02) | H | 0.60(0.01) | 18.46(0.05) | p | 9.92 | 0.18 |
| 0419-487 | 7459( 175) | 7.58(0.02) | H | 0.44(0.023 | 20.13(0.54) | p | 12.55 | 0.83 |
| 0423+044 | 5140( 110) | 8.00(0.00) | H | 0.58(0.00) | 20.72(1.66) | p | 14.75 | 4.96 |
| 0423+120 | 6168( 128) | 8.11(0.06) | He | 0.65(0.04) | 17.36(0.75) | p | 14.12 | 3.06 |
| 0426+588 | 7179( 182) | 8.10(0.02) | He | 0.69(0.02) | 5.54(0.02) | p | 13.54 | 2.02 |
| 0431-279 | 5330( 146) | 8.00(0.00) | H | 0.58(0.00) | 25.15(0.52) | s | 14.59 | 4 |
| 0431-360 | 5153( 154) | 8.00(0.00) | H | 0.58(0.00) | 25.67(0.62) | s | 14.74 | 4.89 |
| 0433+270 | 5629( 98) | 8.06(0.04) | H | 0.62(0.03) | 17.69(0.40) | p | 14.44 | 3.36 |
| 0435-088 | 6368( 117) | 7.91(0.04) | He | 0.53(0.02) | 9.51(0.23) | p | 13.72 | 1.85 |
| 0454+620 | 11347( 967) | 8.67(0.12) | H | 1.02(0.07) | 21.60(0.90) | s | 12.39 | 1.3 |
| 0503-174 | 5316( 88) | 7.62(0.15) | H | 0.39(0.08) | 21.93(1.92) | p | 14.12 | 2.25 |
| 0511+079 | 6661( 133) | 8.22(0.20) | H | 0.73(0.13) | 20.27(0.60) | p | 13.94 | 2.81 |
| 0532+414 | 7760( 112) | 7.71(0.06) | H | 0.45(0.03) | 24.35(0.48) | s | 12.55 | 0.89 |
| 0541+260 | 4460(393) | 8.00(0.00) | He | 0.58(0.00) | 20.40(3.20) | s | 15.37 | 7.2 |
| 0548-001 | 6133( 104) | 8.17(0.04) | He | 0.69(0.03) | 11.07(0.34) | p | 14.24 | 3.63 |
| 0552-041 | 5182( 81) | 8.37(0.02) | H | 0.82(0.01) | 6.41(0.03) | p | 15.28 | 7.31 |
| 0553+053 | 5785( 105) | 8.22(0.04) | H | 0.73(0.03) | 7.99(0.23) | p | 14.56 | 4.12 |
| 0618+067 | 5940( 120) | 8.27(0.14) | H | 0.76(0.09) | 22.62(2.15) | p | 14.52 | 4.03 |
| 0620-402 | 5919( 278) | 8.00(0.00) | He | 0.59(0.00) | 21.50(0.81) | p | 14.12 | 2.39 |
| 0628-020 | 6913( 62) | 8.10(0.16) | H | 0.66(0.10) | 20.48(0.46) | p | 13.6 | 2.04 |
| 0642-166 | 25193( 37) | 8.56(0.01) | H | 0.97(0.01) | 2.63(0.01) | p | 8.68 | 0.11 |
| 0644+025 | 7335( 182) | 8.56(0.12) | H | 0.95(0.07) | 18.45(1.87) | p | 14.08 | 3.48 |
| 0644+375 | 22288( 341) | 8.10(0.05) | H | 0.69(0.03) | 15.26(0.42) | p | 8.45 | 0.07 |
| 0649+639 | 6231( 104) | 8.43(0.23) | H | 0.87(0.15) | 20.98(1.75) | s | 14.56 | 4.45 |
| 0651-398A | 7214( 135) | 8.00(0.00) | H | 0.55(0.00) | 25.1(4.3) | s | 13.14 | 1.3 |
| 0651-39BB | 7223( 219) | 8.00(0.00) | H | 0.58(0.00) | 25.1(4.3) | s | 13.71 | 1.3 |
| 0655-390 | 6312( 211) | 7.68(0.19) | H | 0.58(0.00) | 17.28(0.30) | s | 13.81 | 1.94 |
| 0657+320 | 4999( 15) | 8.09(0.03) | H | 0.64(0.02) | 18.69(0.31) | p | 15.01 | 6.7 |
| 0659-063 | 6628( 105) | 8.33(0.10) | H | 0.80(0.06) | 12.35(4.00) | p | 14.14 | 3.41 |
| 0706+377 | 6591( 140) | 7.95(0.10) | H | 0.56(0.05) | 24.27(1.41) | p | 13.58 | 1.69 |
| 0708-670 | 5097( 116) | 8.00(0.00) | He | 0.57(0.00) | 17.03(0.30) | s | 14.79 | 5.67 |
| 0727+482.1 | 4934( 81) | 7.89(0.02) | H | 0.599(0.01 | 11.11(0.12) | p | 14.79 | 5.15 |
| 0727+482.2 | 4926( 85) | 8.10(0.01) | H | 0.634(0.01 | 11.11(0.12) | p | 15.09 | 6.57 |
| 0728+642 | 5135( 71) | 8.00(0.00) | H | 0.58(0.00) | 18.39(0.24) | s | 14.75 | 4.99 |
| 0736+053 | 7876( 433) | 7.92(0.00) | He | 0.63(0.00) | 3.51(0.01) | p | 13.02 | 1.37 |
| 0738-172 | 7654( 226) | 7.92(0.04) | He | 0.62(0.02) | 9.10(0.05) | p | 13.12 | 1.44 |
| 0743-336 | 4462( 85) | 7.96(0.01) | H | 0.56(0.01) | 15.21(0.12) | p | 15.32 | 7.34 |
| 0747+073.1 | 4354( 40) | 7.84(0.04) | H | 0.50(0.02) | 18.28(0.23) | p | 15.28 | 6.3 |
| 0747+073.2 | 4723( 37) | 7.95(0.03) | H | 0.55(0.02) | 18.28(0.23) | p | 15.06 | 6.3 |
| 0749+426 | 4585( 74) | 8.00(0.00) | H | 0.58(0.00) | 24.29(0.88) | s | 15.25 | 7.29 |
| 0751-252 | 5085( 139) | 8.01(0.03) | H | 0.59(0.02) | 18.17(0.26) | p | 14.81 | 5.31 |
| 0752-676 | 5735( 103) | 8.23(0.08) | H | 0.74(0.05) | 7.90(0.08) | p | 14.62 | 4.32 |
| 0805+356 | 9233( 40) | 8.51(0.05) | H | 0.93(0.03) | 23.12(0.37) | s | 12.98 | 1.9 |
| 0806-661 | 9996( 390) | 7.76(0.05) | He | 0.88(0.03) | 19.17(0.61) | p | 11.74 | 0.62 |

Table 3—Continued

| WD Num. | $T_{eff}(K)(\sigma)$ | log g ($\sigma$) | Comp | $M_\odot$ ($\sigma$) | Dist. (pc)($\sigma$) | Method | $M_{bol}$ | Age(Gyr) |
|---|---|---|---|---|---|---|---|---|
| 0810+489 | 6663( 141) | 7.96(0.00) | He | 0.57(0.00) | 18.3(0.19) | s | 13.62 | 1.87 |
| 0816-310 | 6464( 370) | 7.97(0.00) | He | 0.57(0.00) | 24.66(0.73) | s | 13.77 | 2.02 |
| 0821-669 | 5088( 137) | 8.12(0.02) | H | 0.66(0.01) | 10.65(0.12) | p | 14.98 | 6 |
| 0827+328 | 7210( 178) | 8.31(0.11) | H | 0.80(0.07) | 22.27(0.19) | p | 13.75 | 2.77 |
| 0839-327 | 8962( 323) | 7.44(0.14) | H | 0.33(0.04) | 8.80(0.15) | p | 11.54 | 0.44 |
| 0840-136 | 4874( 177) | 8.00(0.00) | He | 0.57(0.00) | 13.9(0.80) | s | 14.99 | 6.31 |
| 0856+331 | 9890( 244) | 8.47(0.08) | He | 1.05(0.05) | 20.49(1.43) | p | 13.04 | 2.09 |
| 0912+536 | 7236( 195) | 8.20(0.03) | He | 0.75(0.02) | 10.28(0.20) | p | 13.67 | 2.45 |
| 0946+534 | 8166( 245) | 8.06(0.11) | He | 0.74(0.08) | 22.99(1.85) | p | 13.12 | 1.59 |
| 0955+247 | 8630( 229) | 8.01(0.15) | H | 0.60(0.09) | 24.45(2.69) | p | 12.48 | 0.91 |
| 0959+149 | 7001( 500) | 7.94(0.00) | H | 0.56(0.00) | 25.05(0.69) | s | 13.3 | 1.44 |
| 1008+290 | 4562( 161) | 8.17(0.01) | He | 0.68(0.01) | 14.82(0.10) | s | 15.52 | 7.58 |
| 1009-184 | 6037( 364) | 8.01(0.03) | He | 0.59(0.02) | 18.00(0.28) | p | 14.09 | 2.63 |
| 1019+637 | 6743( 155) | 7.93(0.09) | H | 0.55(0.04) | 16.34(0.96) | p | 13.45 | 1.57 |
| 1033+714 | 4727( 87) | 8.00(0.00) | H | 0.58(0.00) | 15.15(0.00) | s | 15.12 | 6.81 |
| 1036-204 | 4694( 124) | 8.05(0.02) | He | 0.60(0.01) | 14.29(0.13) | p | 15.22 | 7.13 |
| 1043-188 | 5788( 85) | 7.93(0.19) | He | 0.53(0.11) | 17.57(2.01) | p | 14.17 | 2.62 |
| 1055-072 | 7421( 200) | 8.32(0.06) | He | 0.85(0.04) | 12.15(0.52) | p | 13.74 | 2.93 |
| 1105-048 | 15141( 88) | 7.85(0.02) | H | 0.54(0.01) | 17.33(3.75) | p | 9.78 | 0.16 |
| 1116-470 | 5801( 176) | 8.00(0.00) | He | 0.57(0.00) | 18.56(0.09) | s | 14.21 | 2.94 |
| 1121+216 | 7435( 185) | 8.08(0.05) | H | 0.65(0.03) | 13.44(0.51) | p | 13.25 | 1.65 |
| 1132-325 | 5000( 500) | 8.00(0.00) | H | 0.58(0.00) | 9.56(0.03) | p | 14.87 | 5.69 |
| 1134+300 | 22469( 342) | 8.56(0.05) | H | 0.97(0.03) | 15.63(0.84) | p | 9.19 | 0.17 |
| 1142-645 | 7972( 219) | 7.88(0.02) | He | 0.61(0.01) | 4.63(0.02) | p | 12.92 | 1.29 |
| 1145-451 | 6101( 150) | 7.99(0.24) | H | 0.58(0.12) | 22.94(2.08) | s | 13.97 | 2.13 |
| 1148+687 | 6620( 13) | 8.24(0.02) | H | 0.75(0.01) | 16.96(0.11) | s | 14 | 2.96 |
| 1149-272 | 6119( 194) | 7.99(0.00) | H | 0.58(0.00) | 25.39(0.49) | s | 13.96 | 2.11 |
| 1202-232 | 8726( 126) | 7.73(0.05) | H | 0.46(0.02) | 10.83(0.11) | p | 12.05 | 0.67 |
| 1208+076 | 5270(14) | 8.00(000) | H | 0.58(0.00) | 24.7(2.80) | s | 14.64 | 4.3 |
| 1208+576 | 5870( 111) | 7.96(0.15) | H | 0.57(0.08) | 20.45(1.92) | p | 14.11 | 2.38 |
| 1223-659 | 7603( 108) | 7.59(0.05) | H | 0.39(0.02) | 16.25(0.31) | p | 12.48 | 0.8 |
| 1236-495 | 11309( 475) | 8.47(0.18) | H | 0.90(0.11) | 16.39(2.53) | p | 12.02 | 0.92 |
| 1241-798 | 9419( 242) | 7.73(0.00) | H | 0.47(0.00) | 22.60(0.41) | s | 11.72 | 0.55 |
| 1257+037 | 5616( 100) | 8.19(0.09) | H | 0.71(0.06) | 16.58(1.05) | p | 14.65 | 4.35 |
| 1309+853 | 5440( 98) | 8.20(0.03) | H | 0.71(0.02) | 16.50(0.27) | p | 14.8 | 4.99 |
| 1310+583 | 10433( 156) | 7.89(0.05) | H | 0.55(0.02) | 24.90(1.00) | s | 11.48 | 0.49 |
| 1310-472 | 4158( 72) | 8.09(0.06) | H | 0.64(0.04) | 15.04(0.54) | p | 15.81 | 8.87 |
| 1315-781 | 5619( 193) | 8.17(0.02) | H | 0.70(0.01) | 19.18(0.35) | p | 14.61 | 4.19 |
| 1327-083 | 14571( 235) | 7.99(0.04) | H | 0.53(0.08) | 16.25(0.54) | p | 10.15 | 0.21 |
| 1334+039 | 4971( 83) | 7.94(0.05) | H | 0.55(0.03) | 8.24(0.23) | p | 14.82 | 5.38 |
| 1336+052 | 4180(46) | 8.00(0.00) | H | 0.58(0.00) | 13.20(1.50) | s | 15.65 | 8.53 |
| 1337+705 | 20464( 103) | 7.90(0.01) | H | 0.58(0.01) | 24.80(1.65) | p | 8.52 | 0.06 |
| 1338+052 | 4360( 50) | 8.00(0.00) | H | 0.58(0.00) | 14.37(0.13) | s | 15.47 | 8.01 |
| 1339-340 | 5258( 166) | 8.00(0.00) | H | 0.58(0.00) | 21.03(0.47) | s | 14.65 | 4.36 |
| 1344+106 | 7060( 168) | 8.02(0.10) | H | 0.61(0.06) | 20.04(1.45) | p | 13.39 | 1.59 |
| 1344+572 | 13960( 294) | 8.05(0.05) | H | 0.64(0.03) | 24.49(0.34) | s | 10.44 | 0.28 |
| 1345+238 | 4581( 85) | 7.77(0.04) | H | 0.46(0.02) | 12.06(0.32) | p | 14.97 | 5.31 |
| 1350-090 | 9458( 136) | 7.86(0.05) | H | 0.53(0.02) | 25.30(1.00) | s | 11.87 | 0.62 |
| 1401+457 | 2670(1500) | 8.00(0.00) | H | 0.57(0.00) | 24.00(10.0) | s | 17.6 | 11.98 |

Table 3—Continued

| WD Num. | $T_{eff}(K)(\sigma)$ | log g ($\sigma$) | Comp | $M_\odot$ ($\sigma$) | Dist. (pc)($\sigma$) | Method | $M_{bol}$ | Age(Gyr) |
|---|---|---|---|---|---|---|---|---|
| 1425-811 | 12128( 182) | 7.98(0.04) | H | 0.60(0.02) | 22.73(0.30) | s | 10.94 | 0.36 |
| 1436-781 | 6271( 200) | 8.06(0.03) | H | 0.63(0.02) | 24.65(0.49) | p | 13.97 | 2.36 |
| 1443+256 | 5190(52) | 8.00(0.00) | H | 0.58(0.00) | 17.50(2.00) | s | 14.7 | 4.7 |
| 1444-174 | 5205( 14) | 8.49(0.08) | H | 0.90(0.05) | 14.49(0.84) | p | 15.44 | 8.08 |
| 1524+297 | 5110(109) | 8.00(0.00) | H | 0.58(0.00) | 22.40(2.60) | s | 14.77 | 5.11 |
| 1532+129 | 6000(400) | 7.88(0.00) | H | 0.53(0.00) | 22.40(2.60) | s | 12.92 | 1.14 |
| 1544-377 | 10348( 151) | 7.70(0.04) | H | 0.46(0.02) | 15.35(0.09) | p | 11.26 | 0.42 |
| 1609+135 | 9041( 259) | 8.47(0.10) | H | 0.90(0.06) | 18.35(1.58) | p | 13 | 1.91 |
| 1620-391 | 25985( 369) | 7.96(0.04) | H | 0.68(0.02) | 12.79(0.06) | p | 7.55 | 0.02 |
| 1625+093 | 7271( 146) | 8.73(0.17) | H | 1.04(0.10) | 23.36(2.02) | p | 14.44 | 3.65 |
| 1626+368 | 8493( 320) | 7.76(0.05) | He | 0.58(0.03) | 15.95(0.51) | p | 12.57 | 1.02 |
| 1630+089 | 5630( 79) | 8.00(0.04) | H | 0.59(0.02) | 13.29(0.16) | s | 14.34 | 2.92 |
| 1632+177 | 9986( 145) | 7.49(0.04) | H | 0.35(0.01) | 21.41(0.23) | s | 11.13 | 0.35 |
| 1633+433 | 6609( 148) | 8.11(0.06) | H | 0.66(0.04) | 15.11(0.68) | p | 13.81 | 2.31 |
| 1633+572 | 5958( 116) | 8.00(0.06) | He | 0.57(0.04) | 14.45(0.52) | p | 14.12 | 2.62 |
| 1639+537 | 7512( 210) | 8.02(0.11) | H | 0.61(0.07) | 21.10(1.56) | p | 13.11 | 1.36 |
| 1647+591 | 12562( 201) | 8.22(0.05) | H | 0.75(0.03) | 10.99(0.28) | p | 11.17 | 0.5 |
| 1655+215 | 9070( 241) | 7.59(0.11) | H | 0.40(0.05) | 23.26(1.68) | p | 11.7 | 0.52 |
| 1658+440 | 30510( 503) | 9.36(0.09) | H | 1.33(0.03) | 27.43(0.86) | s | 9.52 | 0.32 |
| 1705+030 | 6585( 207) | 8.15(0.14) | He | 0.69(0.09) | 17.54(1.66) | p | 13.92 | 2.54 |
| 1748+708 | 5570( 107) | 8.34(0.02) | He | 0.81(0.01) | 6.07(0.09) | p | 14.92 | 5.6 |
| 1756+143 | 5167( 175) | 8.00(0.00) | H | 0.58(0.00) | 20.50(1.20) | s | 14.72 | 4.82 |
| 1756+827 | 7215( 146) | 7.88(0.07) | H | 0.53(0.03) | 15.65(0.71) | p | 13.1 | 1.27 |
| 1814+134 | 5251( 155) | 8.15(0.03) | H | 0.68(0.02) | 14.22(0.24) | p | 14.88 | 5.48 |
| 1817-598 | 4960( 145) | 8.00(0.00) | H | 0.58(0.00) | 24.88(2.10) | p | 14.91 | 5.85 |
| 1820+609 | 4921( 11) | 7.96(0.09) | He | 0.56(0.05) | 12.79(0.67) | p | 14.89 | 5.6 |
| 1829+547 | 6346( 136) | 8.48(0.11) | H | 0.90(0.07) | 14.97(1.25) | p | 14.57 | 4.65 |
| 1840+042 | 9030( 130) | 8.16(0.05) | H | 0.70(0.03) | 24.88(2.10) | p | 12.52 | 1.17 |
| 1848-689 | 4800( 200) | 8.00(0.00) | H | 0.58(0.00) | 24.96(1.35) | p | 15.05 | 6.52 |
| 1900+705 | 11517( 460) | 8.46(0.05) | H | 0.90(0.03) | 12.99(0.39) | p | 11.93 | 0.86 |
| 1911+536 | 17670( 100) | 8.32(0.05) | H | 0.82(0.03) | 22.04(0.44) | s | 9.83 | 0.22 |
| 1912+143 | 6941( 113) | 8.55(0.21) | H | 0.94(0.12) | 19.97(1.88) | s | 14.3 | 3.88 |
| 1917+386 | 6460( 143) | 8.26(0.05) | He | 0.76(0.03) | 11.70(0.47) | p | 14.17 | 3.22 |
| 1917-077 | 10174( 357) | 7.83(0.04) | He | 0.62(0.02) | 10.31(0.23) | p | 11.77 | 0.65 |
| 1919+145 | 15280( 246) | 8.21(0.04) | H | 0.75(0.03) | 19.80(2.16) | p | 10.29 | 0.29 |
| 1935+276 | 12514( 195) | 7.97(0.05) | H | 0.59(0.02) | 17.95(0.93) | p | 10.8 | 0.33 |
| 1953-011 | 7871( 202) | 8.07(0.04) | H | 0.64(0.03) | 11.39(0.38) | p | 12.98 | 1.36 |
| 2002-110 | 4675( 88) | 8.23(0.02) | H | 0.73(0.01) | 17.33(0.24) | p | 15.51 | 8.08 |
| 2007-303 | 16147( 233) | 7.98(0.04) | H | 0.61(0.02) | 16.37(1.21) | p | 9.69 | 0.15 |
| 2008-600 | 4905( 213) | 7.77(0.02) | H+He | 0.44(0.01) | 16.55(0.24) | p | 14.67 | 3.24 |
| 2008-799 | 5800( 160) | 7.97(0.06) | H | 0.57(0.03) | 22.37(0.95) | p | 14.17 | 2.53 |
| 2011+065 | 6401( 150) | 8.11(0.06) | H | 0.66(0.04) | 24.96(0.84) | p | 13.95 | 2.49 |
| 2032+248 | 20704( 332) | 8.02(0.05) | H | 0.64(0.03) | 14.93(0.42) | p | 8.65 | 0.07 |
| 2039-202 | 20163( 300) | 7.98(0.04) | H | 0.62(0.02) | 21.16(1.54) | p | 8.7 | 0.07 |
| 2039-682 | 17105( 298) | 8.59(0.05) | H | 0.98(0.03) | 18.60(0.43) | s | 10.44 | 0.4 |
| 2040-392 | 10987( 163) | 7.88(0.04) | H | 0.54(0.02) | 22.63(0.50) | p | 11.24 | 0.42 |
| 2047+372 | 14712( 286) | 8.31(0.04) | H | 0.81(0.03) | 17.28(0.20) | p | 10.62 | 0.37 |
| 2048+263 | 5073( 76) | 7.25(0.13) | H | 0.25(0.04) | 20.08(1.37) | p | 13.91 | 1.67 |
| 2048-250 | 7613( 213) | 8.18(0.31) | H | 0.71(0.20) | 24.56(2.26) | s | 13.3 | 1.94 |

Table 3—Continued

| WD Num. | $T_{eff}(K)(\sigma)$ | log g ($\sigma$) | Comp | $M_\odot$ ($\sigma$) | Dist. (pc)($\sigma$) | Method | $M_{bol}$ | Age(Gyr) |
|---|---|---|---|---|---|---|---|---|
| 2054-050 | 4491( 42) | 7.84(0.10) | H | 0.50(0.05) | 16.80(0.88) | p | 15.14 | 5.91 |
| 2058+550 | 4530( 59) | 8.00(0.00) | H | 0.58(0.00) | 22.60(2.50) | s | 15.3 | 7.48 |
| 2058+342A | 12220( 423) | 9.09(0.02) | He | 1.23(0.01) | 25.65(0.40) | s | 12.9 | 1.7 |
| 2058+342B | 3875(400) | 9.00(0.01) | He | 1.19(0.05) | 20.13(1.70) | s | 17.69 | 1.19 |
| 2105-820 | 10034( 306) | 7.98(0.21) | H | 0.59(0.10) | 17.06(2.56) | p | 11.78 | 0.59 |
| 2111+072 | 6471( 66) | 8.18(0.15) | H | 0.70(0.10) | 24.33(2.25) | p | 14 | 2.78 |
| 2117+539 | 14684( 239) | 7.91(0.05) | H | 0.57(0.02) | 19.72(2.88) | p | 10 | 0.19 |
| 2118-388 | 5244( 102) | 8.00(0.00) | H | 0.58(0.00) | 22.43(0.37) | s | 14.66 | 4.43 |
| 2119+040 | 5150( 50) | 8.00(0.00) | H | 0.58(0.00) | 23.02(0.20) | s | 14.74 | 4.91 |
| 2126+734A | 16104( 237) | 7.97(0.04) | H | 0.60(0.02) | 21.23(1.08) | p | 9.68 | 0.15 |
| 2126+734B | 6000(300) | 8.39(0.04) | H | 0.84(0.03) | 21.23(1.08) | p | 14.67 | 4.58 |
| 2133-135 | 10131( 158) | 7.76(0.01) | H | 0.48(0.00) | 23.81(0.22) | s | 11.43 | 0.47 |
| 2138-332 | 7400( 388) | 8.09(0.04) | He | 0.70(0.02) | 15.63(0.34) | p | 13.44 | 1.87 |
| 2140-078 | 4830(515) | 8.00(0.00) | H | 0.58(0.00) | 24.20(3.20) | s | 15.02 | 6.4 |
| 2140+207 | 8256( 255) | 7.61(0.07) | He | 0.48(0.04) | 12.52(0.50) | p | 12.17 | 0.82 |
| 2149+021 | 18170( 266) | 8.01(0.04) | H | 0.63(0.02) | 24.51(1.50) | p | 9.21 | 0.11 |
| 2151-015 | 8504( 300) | 8.15(0.06) | H | 0.69(0.04) | 19.85(0.40) | s | 12.77 | 1.35 |
| 2154-512 | 7194( 92) | 7.95(0.08) | He | 0.60(0.04) | 15.87(0.71) | p | 13.34 | 1.63 |
| 2159-754 | 8862( 130) | 8.23(0.06) | H | 0.75(0.04) | 25.16(0.56) | s | 12.72 | 1.43 |
| 2210+565 | 16790( 166) | 8.15(0.15) | H | 0.71(0.09) | 19.54(0.34) | s | 9.78 | 0.19 |
| 2211-392 | 6151( 135) | 8.33(0.07) | H | 0.80(0.05) | 18.69(0.91) | p | 14.46 | 4.01 |
| 2215+386 | 4993( 192) | 8.00(0.00) | H | 0.58(0.00) | 25.13(3.16) | p | 14.88 | 5.72 |
| 2226-754 | 4284( 138) | 8.00(0.00) | H | 0.58(0.00) | 13.43(0.36) | s | 15.55 | 8.24 |
| 2226-755 | 4211( 143) | 8.00(0.00) | H | 0.58(0.00) | 14.42(0.41) | s | 15.62 | 8.45 |
| 2246+223 | 10212( 326) | 8.33(0.09) | H | 0.81(0.06) | 19.05(1.49) | p | 12.25 | 1.07 |
| 2248+293 | 5592( 98) | 7.55(0.16) | H | 0.36(0.08) | 20.92(1.84) | p | 13.8 | 1.61 |
| 2251-070 | 4000( 8) | 8.01(0.06) | He | 0.58(0.03) | 8.52(0.07) | p | 15.87 | 8.48 |
| 2253+054 | 6241( 154) | 8.60(0.24) | H | 0.97(0.14) | 24.46(1.27) | p | 14.85 | 4.83 |
| 2307+548 | 5700(86) | 8.00(0.00) | H | 0.59(0.00) | 16.20(0.70) | s | 14.29 | 2.79 |
| 2307-691 | 9837(2000) | 7.94(0.00) | H | 0.57(0.00) | 20.94(0.94) | p | 11.82 | 0.6 |
| 2311-068 | 7442( 190) | 7.98(0.20) | H | 0.58(0.10) | 25.13(2.97) | p | 13.09 | 1.28 |
| 2322+137 | 5179( 71) | 7.54(0.08) | H | 0.35(0.04) | 22.27(0.99) | p | 14.14 | 2.09 |
| 2326+049 | 11956( 187) | 8.01(0.05) | H | 0.61(0.03) | 13.62(0.74) | p | 11.05 | 0.38 |
| 2336-079 | 10678( 317) | 8.05(0.03) | H | 0.64(0.02) | 15.94(0.43) | p | 11.62 | 0.58 |
| 2341+322 | 13111( 198) | 7.92(0.04) | H | 0.56(0.053) | 17.59(0.55) | p | 10.52 | 0.27 |
| 2345+027 | 4900(47) | 8.00(0.00) | H | 0.58(0.00) | 22.70(3.60) | s | 14.96 | 6.17 |
| 2347+292 | 5805( 110) | 7.83(0.15) | H | 0.50(0.08) | 22.40(2.06) | p | 13.99 | 2.17 |
| 2351-335 | 8762( 128) | 7.71(0.06) | H | 0.45(0.03) | 22.86(0.75) | p | 12.01 | 0.65 |
| 2359-434 | 8645( 123) | 8.04(0.05) | H | 0.62(0.03) | 8.17(0.07) | p | 12.52 | 0.98 |